\documentclass[fleqn,usenatbib]{mnras}

\usepackage{newtxtext,newtxmath}

\usepackage[T1]{fontenc}

\DeclareRobustCommand{\VAN}[3]{#2}
\let\VANthebibliography\thebibliography
\def\thebibliography{\DeclareRobustCommand{\VAN}[3]{##3}\VANthebibliography}

\usepackage{graphicx}	
\usepackage{amsmath}	

\title[Drivers of the Gas-Phase Metallicity]{Both Stellar Mass and Gravitational Potential Shape the Gas-Phase Metallicity}

\author[M. Koller et al.]{
Maria Koller$^{1,2}$\thanks{E-mail:mk2264@cam.ac.uk},
Roberto Maiolino $^{1,2,3}$,
William M. Baker $^{4}$
\\
$^{1}$Kavli Institute for Cosmology, University of Cambridge, Madingley Road, Cambridge, CB3 OHA, UK\\
$^{2}$Cavendish Laboratory, University of Cambridge, 19 JJ Thomson Avenue, Cambridge CB3 0HE, UK\\
$^{3}$Department of Physics and Astronomy, University College London, Gower Street, London WC1E 6BT, UK \\
$^{4}$DARK, Niels Bohr Institute, University of Copenhagen, Jagtvej 155A, DK-2200 Copenhagen, Denmark \\
}

\date{Accepted XXX. Received YYY; in original form ZZZ}

\pubyear{\the\year{}}

\begin{document}
\label{firstpage}
\pagerange{\pageref{firstpage}--\pageref{lastpage}}
\maketitle

\begin{abstract}
The relation between metallicity and galaxy mass (the so-called mass-metallicity relation)  is the strongest and most prominent among scaling relations between chemical enrichment and galactic properties. However, it is unclear whether this relation primarily traces metal retention or the integrated production of metals, as past studies have obtained contrasting results. We investigate this issue through an extensive Random Forest and Partial Correlations analysis of spectral cubes of 4,500 galaxies from the MaNGA survey. We find that stellar mass ($\rm M_*$) and baryonic gravitational potential ($\rm \Phi_* = M_*/R_e$) are the two most important quantities determining gas metallicity in galaxies. However, their relative roles strongly depend on the galactocentric radius -- the metallicity within 0.7~$\rm R_e$ depends primarily on the stellar mass, while the metallicity at radii beyond 0.9~$\rm R_e$ depends primarily on the gravitational potential. This finding can be interpreted in terms of metals in the central region ($\rm R\leq 0.7~R_e$) being mostly bound, regardless of the global gravitational potential and, therefore, the metallicity is determined primarily by the cumulative production of metals (hence the integrated star formation history, i.e. $\rm M_*$); by contrast, in the galactic peripheries the retention of metals depends more critically on the gravitational potential, hence the stronger dependence of the metallicity on $\rm \Phi_*$ at large radii. Our finding reconciles apparent discrepancies between previous results. Finally, we find that the Star Formation Rate is the third most important parameter (after $\rm M_*$ and $\rm \Phi_*$) in determining the metallicity, as expected from the Fundamental Metallicity Relation.
\end{abstract}




\section{Introduction}

Galaxy evolution is governed by the regulation of gas content, metallicity, and stellar mass, shaped by gas inflows from the IGM, feedback-driven outflows, and internal recycling \citep{madau_cosmic_2014, lilly_gas_2013, peroux_cosmic_2020}. Metallicity is linked to the time-integrated production of chemical elements; hence, it reflects a galaxy’s star formation history \citep{maiolino_re_2019}. Metallicity is also more broadly connected to the baryon cycle within and around galaxies. Indeed, inflows dilute ISM metals while fueling star formation, whereas stellar winds and supernovae enrich the ISM and CGM \citep{tumlinson_circumgalactic_2017,lilly_gas_2013, peng_haloes_2014}. AGN feedback further influences the ISM and CGM by heating or expelling gas, suppressing star formation \citep{fabian_observational_2012, bourne_recent_2023}. On the other hand, AGN feedback can also trigger star formation within galactic outflows \citep{maiolino_star_2017}. Dust also affects metal depletion, and over time, gas cycles through phases of enrichment and redistribution. Given all these interconnected aspects between chemical enrichment and galaxy evolution, a strong relationship is expected between metallicity and galactic properties.

The first, and most widely studied relation is between gas-phase metallicity and stellar mass, the so-called mass-metallicity relation \citep[MZR; see e.g.,][for a review]{tremonti_origin_2004, kewley_metallicity_2008, sanchez_mass-metallicity_2017, maiolino_re_2019}. This correlation was first defined as a metallicity versus luminosity relation by \cite{lequeux_chemical_1979}, but later shown to be more directly connected with the galaxy mass. Specifically, an increase in stellar mass correlates with an increase in metallicity until the relation flattens out at high stellar masses ($\mathrm{\gtrsim 10^{10} M_{\odot}}$). This relation has been studied for the local Universe and redshifts up to $z \sim 10$ \citep{maiolino_amaze_2008, lamareille_physical_2009, mannucci_fundamental_2010, maier_mass-metallicity_2014, cresci_fundamental_2019, curti_mass-metallicity_2020, curti_chemical_2023, langeroodi_evolution_2023, nakajima_jwst_2023}. The MZR has also been studied at spatially resolved scales \citep{rosales-ortega_new_2012, sanchez_mass-metallicity_2013, sanchez_almeida_fundamental_2019, baker_metallicitys_2022}. This relation has been observed to evolve with redshift: galaxies in the early universe are metal-poorer than local galaxies  \citep{tremonti_origin_2004, savaglio_gemini_2005, maiolino_amaze_2008, gao_what_2018, sanders_mosdef_2021}. Generally, this is attributed to galaxies in the local universe having had more time to assemble and produce metals than at high redshifts. 

The second most widely studied relation is the metallicity's anti-correlation with the star formation rate (SFR), which, together with the stellar mass, gives rise to the so-called Fundamental Metallicity Relation \citep[FMR;][]{mannucci_fundamental_2010, lara-lopez_fundamental_2010, cresci_fundamental_2019, curti_mass-metallicity_2020, pistis_comparative_2024, curti_jades_2024, sarkar_unveiling_2025}. This relation has also been observed to exist at spatially resolved scales \citep{baker_metallicitys_2022, koller_magpi_2024}. Other significant relations of the gas-phase metallicity have been found with the stellar age \citep{lian_massmetallicity_2015, sanchez-menguiano_local_2020}, rotational velocity \citep{garnett_luminosity-metallicity_2002, pilyugin_oxygen_2004, hughes_role_2013}, galaxy size \citep{ellison_clues_2008, sanchez_almeida_origin_2018}, velocity dispersion \citep{li_sdss-iv_2018, cappellari_full_2023}, and gas mass fraction \citep{bothwell_fundamental_2013, barrera-ballesteros_sdss-iv_2018, brown_role_2018}. 

Nevertheless, stellar mass remains the strongest contributor to defining metallicity among those, and the origin of the MZR has long been debated \citep{maiolino_re_2019}. A common explanation is that massive galaxies inhabit a deeper gravitational potential that is more efficient at retaining metals against outflows compared to lower mass galaxies \citep{tremonti_origin_2004, tumlinson_large_2011, chisholm_metal-enriched_2018}. In this sense, both the dynamical mass ($\mathrm{M_{dyn}}$) and gravitational potential, defined as $\mathrm{\Phi_{dyn} = M_{dyn}/R_e}$, would form a stronger correlation with the metallicity than the stellar mass would. Dynamical measurements are much rarer than photometric or spectroscopic estimates; thus, a common approximation of the gravitational potential is $\mathrm{\Phi_* = M_{*}/R_e}$ \citep{deugenio_gas-phase_2018, cappellari_full_2023}. 

\citet{deugenio_gas-phase_2018} investigated the correlation between the metallicity and the stellar mass $\mathrm{M_*}$, gravitational potential $\mathrm{\Phi_*}$, and surface mass density $\mathrm{\Sigma_* = M_*/R_e^2}$ for a volume-limited sample from SDSS DR7. They found that the metallicity more tightly correlates with $\mathrm{\Phi_*}$ rather than $\mathrm{M_*}$ or $\mathrm{\Sigma_*}$. 

\citet{sanchez-menguiano_stellar_2024} used data from the MaNGA pyPipe3D value-added catalogue (VAC) of more than 3000 galaxies to simultaneously investigate the effect of 148 parameters on the gas-phase metallicity measured at $\rm 1R_e$ via random forest regression. Their results confirm that $\mathrm{\Phi_{*}}$, as a proxy for the baryonic gravitation potential, is the dominating property in determining the gas-phase metallicity. They also find that the photometric mass is more important in defining the metallicity than the stellar mass measured via spectroscopy. 

\citet{boardman_tight_2024} likewise used MaNGA data and reported $\mathrm{\Phi_{*}}$ to correlate even more tightly with the gaseous $\rm N/O$ abundance ratio than with the gas-phase metallicity. \citet{boardman_competing_2025} subsequently found both abundance measures to possess secondary correlations with galaxies' positions along the star-forming sequence, analogous to the FMR.

A tight correlation with $\mathrm{\Phi_*}$ has also been found for the stellar metallicity \citep{barone_sami_2018, vaughan_sami_2022, cappellari_full_2023, sanchez-menguiano_more_2024}, consistent with the stellar fundamental metallicity relation reported by \citet{looser_stellar_2024}, who found that $\mathrm{\Phi_*}$ provides a tighter correlation than stellar mass alone.

On the contrary, an investigation done by \citet{baker_stellar_2023} found that the stellar mass $\mathrm{M_*}$, not $\mathrm{\Phi_*}$ or the dynamical measurements $\mathrm{M_{dyn}}$ and $\mathrm{\Phi_{dyn}}$, is the dominating factor in determining the gas-phase metallicity. Their study focused on Partial Correlation Coefficient (PCC) analysis and random forest regression \citep{bluck_are_2020, baker_almaquest_2022} of roughly 1000 galaxies from the MaNGA survey. 

Despite using multi-dimensional regression techniques and machine learning tools, these conflicting results leave the question of the primary driver of galaxy metallicity and the origin of the MZR largely open.

In this paper, we aim to unravel the nature of the MZR and the origin of the conflicting past results by using data from the full MaNGA and SDSS surveys of local star-forming galaxies. We will illustrate that there is no contradiction between previous results and that the different findings originate from the different scales probed in other studies. 

This paper is structured as follows. In section~\ref{sec:data} , we introduce the MaNGA and SDSS survey, the data products used for this analysis, and describe our sample selection and derivation of physical quantities. In section~\ref{sec:analysis} , we present our analysis technique, including Partial Correlation Coefficients (section~\ref{sec:analysis_PCC}) and random forest Regression (section~\ref{sec:analysis_RF}). Our results are shown in section~\ref{sec:results}, and their physical implications are discussed in \ref{sec:discussion}. Finally, we summarise our key results in section~\ref{sec:conclusion}.

Throughout this paper, we assume a \cite{kroupa_variation_2001} initial mass function (IMF) and adopt a flat $\mathrm{\Lambda CDM}$ cosmology with $\mathrm{H_0 = 70.0 km s^{-1} Mpc^{-1}}$, $\mathrm{\Omega_m = 0.3}$, and $\mathrm{\Omega_{\lambda} = 0.7}$.

\section{Data}\label{sec:data}

\subsection{MaNGA}\label{sec:data_manga}

We utilise publicly available integral field spectroscopic (IFS) data from MaNGA (Mapping Nearby Galaxies at Apache Point Observatory, \citealt{bundy_overview_2015}), part of the fourth-generation Sloan Digital Sky Survey \citep[SDSS-IV,][]{blanton_sloan_2017}. We use the final and fully complete data release (DR) 17, released in 2022, for 10,000 galaxies \citep{abdurrouf_seventeenth_2022,sanchez_sdss-iv_2022}.  

The MaNGA survey gathered spatially resolved integral field spectroscopy for $\sim 10,000$ galaxies in the nearby universe ($ z\sim$0.03) using the two BOSS spectrographs of the Sloan 2.5 telescope at Apache Point Observatory \citep{gunn_25_2006}. The spectrograph covers wavelengths over 3600--10300 \r{A} at a resolution of $R\sim$2000. The 17 hexagonally distributed fibre bundles result in a field of view (FoV) that varies in diameter from $12^{\prime\prime}$ to  $32^{\prime\prime}$. Galaxies were selected to follow a roughly flat distribution in $i$-band luminosity, which acts as a proxy for the stellar mass. The sample can be divided into a primary and secondary sample: primary galaxies have uniform radial coverage to radii of 1.5$\rm R_e$ (2/3 of the entire sample), while the secondary sample covers out to 2.5$R_e$ (1/3 of the entire sample). The resulting data has a point spread function (PSF) full-width-half-maximum (FWHM) of $\sim 2.5^{\prime\prime}$ corresponding to a physical resolution of $\sim 1.5 $ kpc for each observed spaxel of size $0.5^{\prime\prime} \times 0.5^{\prime\prime}$. 
More details about the survey's design, data selection, flux calibration and data reduction can be found in \citet{law_observing_2015, yan_sdss-iv_2016, wake_sdss-iv_2017, law_sdss-iv_2021}.

For our analysis, we use both the MaNGA Data Analysis Pipeline \citep[DAP,][]{westfall_data_2019, belfiore_data_2019} and pyPipe3D \citep{sanchez_sdss-iv_2022} data. Resolved data is obtained for emission line fluxes, stellar masses, galactocentric distances, and equivalent widths. Emission line fluxes and equivalent widths for $\mathrm{H\alpha}$, $\mathrm{H\beta}$,  $\mathrm{[OIII]\lambda 5007}$, $\mathrm{[OII]\lambda3728}$, $\mathrm{[NII]\lambda 6584}$ and $\mathrm{[SII]\lambda 6717,6732}$ as well as galactocentric distances of each spaxel are obtained from the spatially resolved DAP MAPS products. The DAP uses a hybrid binning scheme via pPXF \citep[penalised pixel-fitting,][]{cappellari_improving_2017} where the emission lines are computed for individual spaxels and the continuum is Voronoi binned \citep{cappellari_adaptive_2003} to achieve a homogeneous S/N ratio. Effective radii, stellar masses per pixel, integrated stellar masses, and integrated star formation rates via H$\alpha$ are taken from the pyPipe3D catalogue. We also utilise photometric stellar masses from \citep{sanchez_sdss-iv_2022} which were measured via synthetic broad-band photometry obtained from the spectral MaNGA datacubes using the colour-M/L relation from \citet{bell_stellar_2000}. We use redshifts and inclinations obtained from the NASA Sloan Atlas \footnote{\url{https://nsatlas.org/data}}. 

We also use dynamical masses $\mathrm{M_{dyn}}$ and effective velocity dispersions, $\mathrm{\sigma_e}$, within elliptical half-light isophotes provided by \cite{zhu_manga_2023}, which were computed via the axisymmetric Jeans Anisotropic Modelling \citep[JAM,][]{cappellari_measuring_2008, cappellari_efficient_2020}.

\subsection{SDSS}\label{sec:data_sdss}

Our second dataset consists of data from the Sloan Digital Sky Survey (SDSS) DR7 \citep{abazajian_seventh_2009}, which includes spectroscopy for around 930,000 galaxies. We get stellar masses and SFRs from the MPA-JHU catalogue\footnote{\url{https://wwwmpa.mpa-garching.mpg.de/SDSS/DR7/}} \citep{kauffmann_host_2003,brinchmann_physical_2004, salim_uv_2007} and use the provided emission line fluxes for computing gas-phase metallicities. We match this data to a catalogue containing morphological parameters \citep{simard_catalog_2011}.

\subsection{Sample selection}\label{sec:data_selection}

Regarding the SDSS data, we employ a signal-to-noise ratio (SNR) cut  $\rm S/N > 3$ for our lines of interest to ensure accurate emission line detection for calculating gas-phase metallicities. We also limit the sample by excluding highly inclined galaxies by applying a minor-to-major axis ratio cut of $\mathrm{b/a\geq 0.35}$.  Furthermore, due to the small aperture of the SDSS observations, some galaxies which are at a very low redshift might only be probed within their most central region. To reduce aperture effects, we restrict our sample to galaxies for which the spectroscopic fibre radius covers more than $0.5  \mathrm{R_e}$. The calibrations used for computing gas-phase metallicities are based on star-forming regions; thus, we select star-forming galaxies based on the $\mathrm{[NII]/H\alpha}$ vs. $\mathrm{[OIII]/H\beta}$ Baldwin-Philips-Terlevich (BPT) diagnostic diagram \citep{baldwin_classification_1981}, using the empirical line by \citet{kauffmann_host_2003}, which separates star-forming galaxies from AGN/LINER galaxies. Our final SDSS sample consists of 121,880 star-forming galaxies. 

For the MaNGA data, we focus on the spatially resolved data products to make our initial sample selection. Via the resolved DAP MAPS files, we make a selection of spaxels with $\rm S/N>3$ in  $\mathrm{H\alpha}$, $\mathrm{H\beta}$,  $\mathrm{[OIII]\lambda 5007}$, $\mathrm{[OII]\lambda3728}$, $\mathrm{[NII]\lambda 6584}$ and $\mathrm{[SII]\lambda 6717,6732}$, and an equivalent width cut of $\mathrm{EW_{H\alpha}}>6 $\r{A} to avoid contamination by diffused ionised gas. Furthermore, we select star-forming spaxels using the same BPT diagram criterion described above for the SDSS data. After defining our sample of spaxels, we compute integrated properties for stellar mass, star formation rate, and metallicity using the spatially resolved values as outlined below. Finally, an inclination cut is applied to exclude any highly inclined galaxies, and a sigma clipping of $\rm 5\sigma$ is applied to exclude any major outliers, resulting in 4571 galaxies. 

\subsection{Derivation of Physical Quantities}\label{sec:data_deriv_quant}

Stellar masses per pixel, integrated total stellar masses and masses within $\mathrm{1R_e}$ are obtained from pyPipe3D \citep{sanchez_sdss-iv_2022}. We obtain galactocentric distances for each spaxel in units of the effective radius from DAP \citep{belfiore_data_2019, westfall_data_2019}. Via the galactocentric distances, we compute total masses and masses within $\mathrm{1R_e}$ by taking the sum of all spaxels. 

Emission lines are corrected for extinction using the $\mathrm{H\alpha/H\beta}$ Balmer decrement, assuming an intrinsic value of $2.86$ \citep{osterbrock_astrophysics_2006}, via the \citet{calzetti_dust_2000} dust attenuation law with $\mathrm{R_v=3.1}$. 

Star formation rates for each spaxel in the MaNGA data are computed via the conversion law from \citet{kennicutt_star_2012} using the $\mathrm{H\alpha}$ emission line:
\begin{equation}
    \mathrm{\log{(SFR  [M_{\odot} yr^{-1}])} = \log{(L_{H\alpha}} [erg \ s^{-1}]) - 41.27}
\end{equation}

We obtain a total SFR within $1\mathrm{R_e}$ by taking the sum of SFRs of the spaxels which fulfil our $\rm S/N$ and BPT diagram requirements. The global catalogue of pyPipe3D data available online\footnote{\url{https://data.sdss.org/datamodel/files/MANGA_PIPE3D/MANGADRP_VER/PIPE3D_VER/SDSS17Pipe3D.html}} also provides integrated SFRs using only the spaxels compatible with star formation. The global values are computed using the \citet{kennicutt_star_1998} calibration for a Salpeter IMF. To convert to the new calibration using a Kroupa IMF, we use the conversion given in \citet{kennicutt_star_2012}:
\begin{equation}
    \mathrm{SFR_{K2012}} = \mathrm{0.68 \cdot SFR_{K1998}}
\end{equation}

We compared our own derived integrated SFRs with those publicly available and found no significant difference in the results shown below. Therefore, we uniformly use the integrated SFRs from the global catalogue for our results unless stated otherwise.

\begin{table}
    \centering
    \renewcommand{\arraystretch}{2}
    \caption{Diagnostics used for computing gas-phase metallicities as presented in \citet{curti_mass-metallicity_2020}.}
    \begin{tabular}{c|c}
    \hline
    Index & Definition \\
    \hline
     R23  &  $\mathrm{\log\left(  \frac{[OIII]\lambda 5007,4958 + [OII]\lambda 3727}{H\beta}\right)}$\\
     O32 &  $\mathrm{\log\left( \frac{[OIII]\lambda 5007}{[OII]\lambda 3727}\right)}$\\
     R2 & $\mathrm{\log\left( \frac{[OII]\lambda 3727}{H\beta}\right)}$ \\
     R3 & $\mathrm{\log\left( \frac{[OIII]\lambda 5007}{H\beta}\right)}$ \\
     N2 & $\mathrm{\log\left( \frac{[NII]\lambda 6584}{H\alpha}\right)}$\\
     S2 &  $\mathrm{\log\left( \frac{[SII]\lambda 6717,6731}{H\alpha}\right)}$\\
     O3N2 & $\mathrm{\log\left( \frac{[OIII]\lambda 5007 / H\beta}{[NII]\lambda 6584 / H\alpha}\right)}$ \\
     O3S2 & $\mathrm{\log\left( \frac{[OIII]\lambda 5007 / H\beta}{[SII]\lambda 6717,6731 / H\alpha}\right)}$\\
     RS32 & $\mathrm{\log\left( \frac{[OIII]\lambda 5007}{H\beta} + \frac{[SII]\lambda 6717,6731}{H\alpha}\right)}$\\
     \hline 
    \end{tabular}
    \label{tab:met_cals}
\end{table}

To obtain a singular value for the gas-phase metallicities, we compute the metallicities per spaxel using nine different strong-line calibrations (see Table~\ref{tab:met_cals}) and take the average over all spaxels from our selection within each galaxy. The diagnostics $\mathrm{R23, R2, R3, O32, N2, S2, O3S2, O3N2,}$ and $\mathrm{R3S2}$ are used in combination with the empirical calibrations presented in \cite{curti_new_2017, curti_mass-metallicity_2020}. Using several diagnostics simultaneously decreases the risk of having degeneracies associated with diagnostics that are double values. It also allows for integrating information from several emission lines that may differ across metallicity regimes, and it allows us to constrain the resulting abundances more precisely (see \citealt{curti_mass-metallicity_2020} for more information). We take the average metallicity value across all selected spaxels within $1\mathrm{R_e}$ to obtain a single value for the gas-phase metallicity per galaxy. We also repeated our analysis using the
median and found no significant difference in the results.

\section{Data Analysis Tools}\label{sec:analysis}

\subsection{Partial Correlation Coefficients}\label{sec:analysis_PCC}

To investigate the direct relationships between variables while accounting for the influence of others, we perform a Partial Correlation Coefficient (PCC) analysis, following a methodology similar to that described in \citet{baker_metallicitys_2022} and \citet{baker_stellar_2023}. This approach allows us to quantify the strength of the association between two variables while statistically removing the effect of other variables. By doing so, we can isolate genuine correlations and distinguish them from those that arise indirectly through shared dependencies.

The PCC between two variables, $\mathrm{A}$ and $\mathrm{B}$, controlled for a third variable, $\mathrm{C}$, is defined as:
\begin{equation}
    \mathrm{\rho_{A B | C} = \frac{\rho_{AB}-\rho_{AC} \ \rho_{BC}}{\sqrt{1-\rho_{AC}^2} \ \sqrt{1-\rho_{BC}^2}}}.
\end{equation}

where $\mathrm{\rho_{XY}}$ represents the Spearman rank correlation coefficient between variables X and Y. This technique is most meaningful when applied to monotonic relationships.

In our analysis, we conceptualise the three variables within a three-dimensional framework: $\mathrm{SFR}$ (y-axis) plotted against $\mathrm{M_{*}}$ (x-axis), with gas-phase metallicity visualised through a third dimension \citep[e.g., colour-coding, see also][]{bluck_are_2020, piotrowska_towards_2020, baker_almaquest_2022,baker_metallicitys_2022, baker_stellar_2023, baker_different_2024}. Here, metallicity corresponds to variable A, $\mathrm{SFR}$ to B, and $\mathrm{M_{*}}$ to C.

To interpret the PCC results, we adopt the arrow angle formalism introduced by \cite{bluck_how_2020}, which identifies the direction of the steepest average increase in metallicity. The angle $\Theta$, measured from the horizontal axis (three o’clock position), is given by \citep{bluck_how_2020}: 
\begin{equation}\label{eq:arrow_angle}
    \mathrm{\tan(\Theta) = \frac{\rho_{B A| C}}{\rho_{C A | B}}}.
\end{equation}

Therefore, this angle represents the ratio of the partial correlation between the y-axis and the colour-coded (z-axis) variable to the partial correlation between the x-axis and z-axis. The direction of the arrow visually illustrates how the z-axis quantity depends on the x and y variables, i.e. the direction of the average gradient of the 3D surface, while the arrow angle provides a quantitative measure of the relative strength of these dependencies (i.e. the direction of the gradient).

To obtain arrow angle uncertainties, we applied the same method as \cite{baker_metallicitys_2022} and \cite{baker_stellar_2023} and utilised 100 bootstrap random samples and computed their standard deviation.

\subsection{Random Forest}\label{sec:analysis_RF}

While PCCs are good for determining the direction of correlations, they do not apply to a wide variety of relations, as they only work for monotonic relations. Thus, we also employ a random forest Regression \citep{breiman_random_2001} to more accurately determine the relative strengths of the correlations. This machine learning approach has previously been demonstrated to be effective in identifying fundamental relationships among galaxy properties \citep[e.g.][]{baker_metallicitys_2022, bluck_quenching_2022,baker_stellar_2023, sanchez-menguiano_stellar_2024}. 

A random forest algorithm works as follows: it makes predictions by combining the results of many decision trees, each trained on different subsets of the data. Averaging the outputs of these trees improves accuracy and reduces the risk of overfitting to noise in the data. Thus, it is advantageous for our analysis as it allows us to simultaneously investigate several intercorrelated parameters. As was found by \citet{bluck_how_2020, bluck_quenching_2022, baker_stellar_2023}, the random forest Regressor can distinguish between intrinsic dependences and intercorrelated parameters. In our analysis, we utilise the random forest Regressor to determine the relative importance of different parameters driving the gas-phase metallicity. 

We base our implementation of the algorithm on the \texttt{scikit-learn} Python package \citep{pedregosa_scikit-learn_2011}. First, we split the data into a train and test set. For the MaNGA sample, we define our test set to be $30\%$ of the data, while for SDSS, we consider half of the selected sample. The training set consists of the predictor variables (all analysed galaxy parameters except the gas-phase metallicity) paired with the target variable (the gas-phase metallicity). The model is fit on this training data, and the test set is used to evaluate its predictive performance and derive the relative importance of the predictors in determining the target quantity. The galaxy properties of interest are the input predictors of the algorithm, and the gas-phase metallicity acts as our target feature. We rely on mean squared error (MSE) measurements of the training and test samples to determine the model's precision. For example, suppose the MSE of the train sample is significantly less than the MSE of the test sample. In that case, we could be overfitting our data, i.e., including noise in finding the parameter importance. On the contrary, if the train MSE is much larger than the test MSE, we could be underfitting our data, meaning that the model is not exploring the full data space when computing the relative importance. To avoid overfitting and underfitting, we fine-tune the hyperparameters of our random forest regressor and use the following hyperparameters universally for both samples: $\mathrm{n_{estimators}=300, max_{depth}=30, min_{split}=2,}$and $\mathrm{min_{samples\_leaf}=6}$. 

\begin{figure*}
    \centering
    \includegraphics[width=\linewidth]{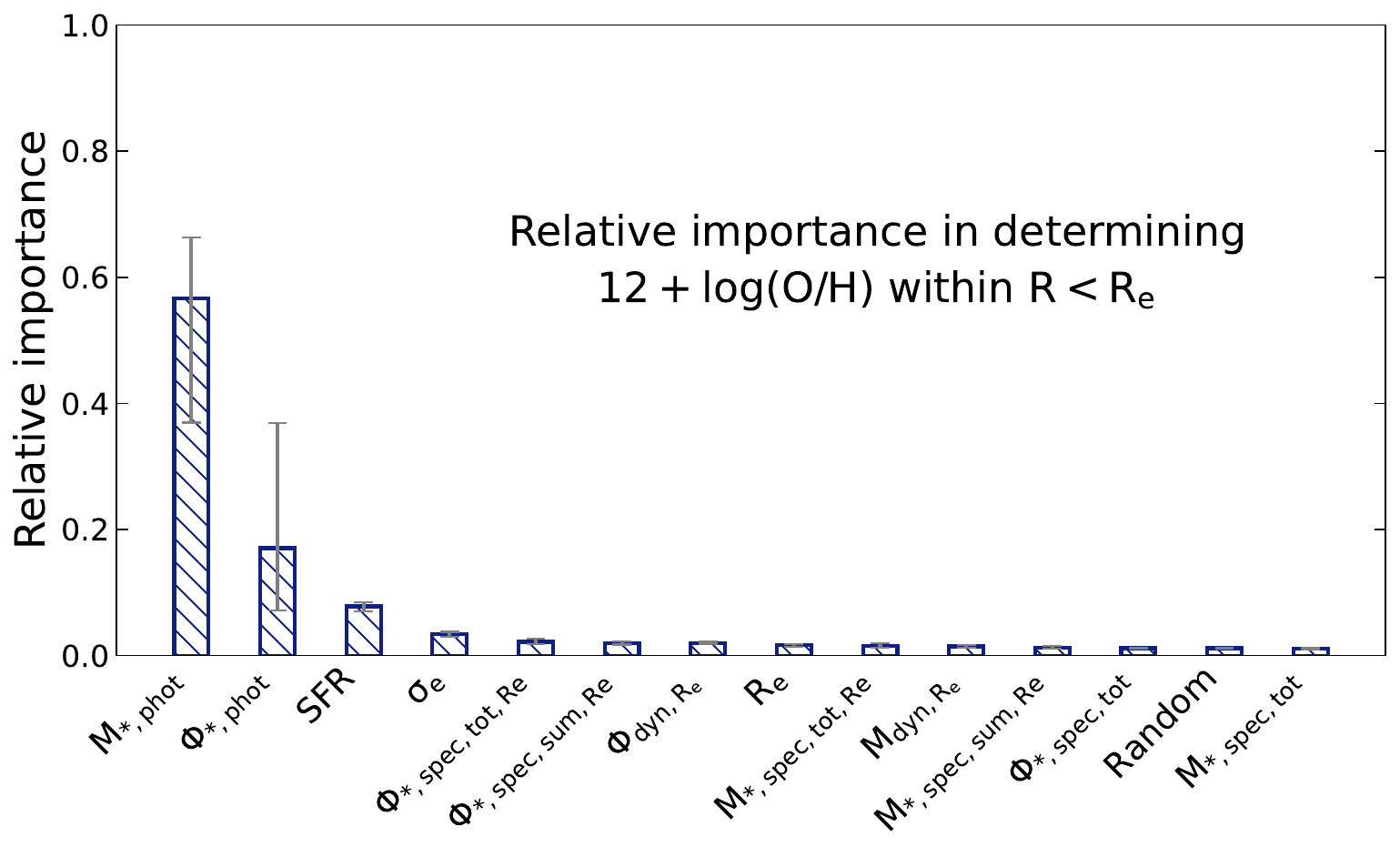}
    \caption{Bar-chart representing the results from random forest regression of the MaNGA sample for our 14 input features in driving the gas-phase metallicity. The error bars represent the 16th and 84th percentiles from 100 bootstrap random samples. The parameters are described in detail in Table~\ref{tab:manga_rf_parameters}. The figure shows that the photometric stellar mass is the main driver of the gas-phase metallicity, with secondary, significant dependences only coming from the baryonic gravitational potential and SFR.}
    \label{fig:manga_rf}
\end{figure*}

\section{Results: What drives the metallicity?}
\label{sec:results}

\subsection{Dominant Features via the Random Forest}\label{sec:results_rf}

\subsubsection{Metallicities within $\rm 1R_e$: the dominance of $\rm M_{*,phot}$}

We start by simultaneously exploring, from our random forest analysis, the relative importance of multiple galactic properties in predicting the average gas-phase metallicity within $\mathrm{1R_e}$. First, we consider our selected MaNGA sample and analyse how different parameters (see table~\ref{tab:manga_rf_parameters} for a description) influence the gas-phase metallicity. We include different stellar mass estimates. The spectroscopic method (denoted as $\mathrm{M_{*,spec}}$) consists of the sum of the spectroscopically determined masses in all spaxels of the MaNGA maps (or all spaxels within $\rm 1R_e$, in which case it is denoted as $\rm M_{*,spec,R_e}$); these are either measured ourselves by taking the sum of spaxels within a certain radius, or taken from the pyPipe3D catalogue. We also use the stellar masses provided in the MPA-JHU and pyPipe3D catalogue, which are determined from the central (single-fibre) SDSS spectroscopic information and corrected with photometry for the extended emission outside the fibre; this is denoted as $\mathrm{M_{*,phot}}$ (although it also has a spectroscopic component). Additionally, similar to previous studies \citep{deugenio_gas-phase_2018, barone_gravitational_2020, vaughan_sami_2022, cappellari_full_2023, baker_stellar_2023,sanchez-menguiano_stellar_2024}, we also include the ``stellar gravitational potential'' inferred via the stellar mass $\mathrm{\Phi_* = M_*/R_e}$, where $\mathrm{M_*}$ is replaced by the different mass estimates, in our analysis. This can be considered a proxy for the total gravitational potential in cases (and in apertures) where stars dominate the gravitational potential. We also include in the random forest analysis the $\mathrm{SFR}$ and effective radius $\mathrm{R_e}$. The former needs to be taken into account due to the metallicity's secondary dependence, defined as the Fundamental Metallicity Relation \citep[FMR, ][]{mannucci_fundamental_2010}. From \citet{zhu_manga_2023} we include several dynamically inferred properties, namely the effective velocity dispersion $\mathrm{\sigma_e}$ and dynamical mass within the effective radius $\mathrm{M_{dyn, Re}}$, with which we also compute the dynamically inferred gravitational potential within $\rm 1 R_e$, $\mathrm{\Phi_{dyn, Re} = M_{dyn, Re}/R_e}$. Lastly, we include a uniform random variable as a control measure.

\begin{figure*}
    \centering
    \includegraphics[width=\linewidth]{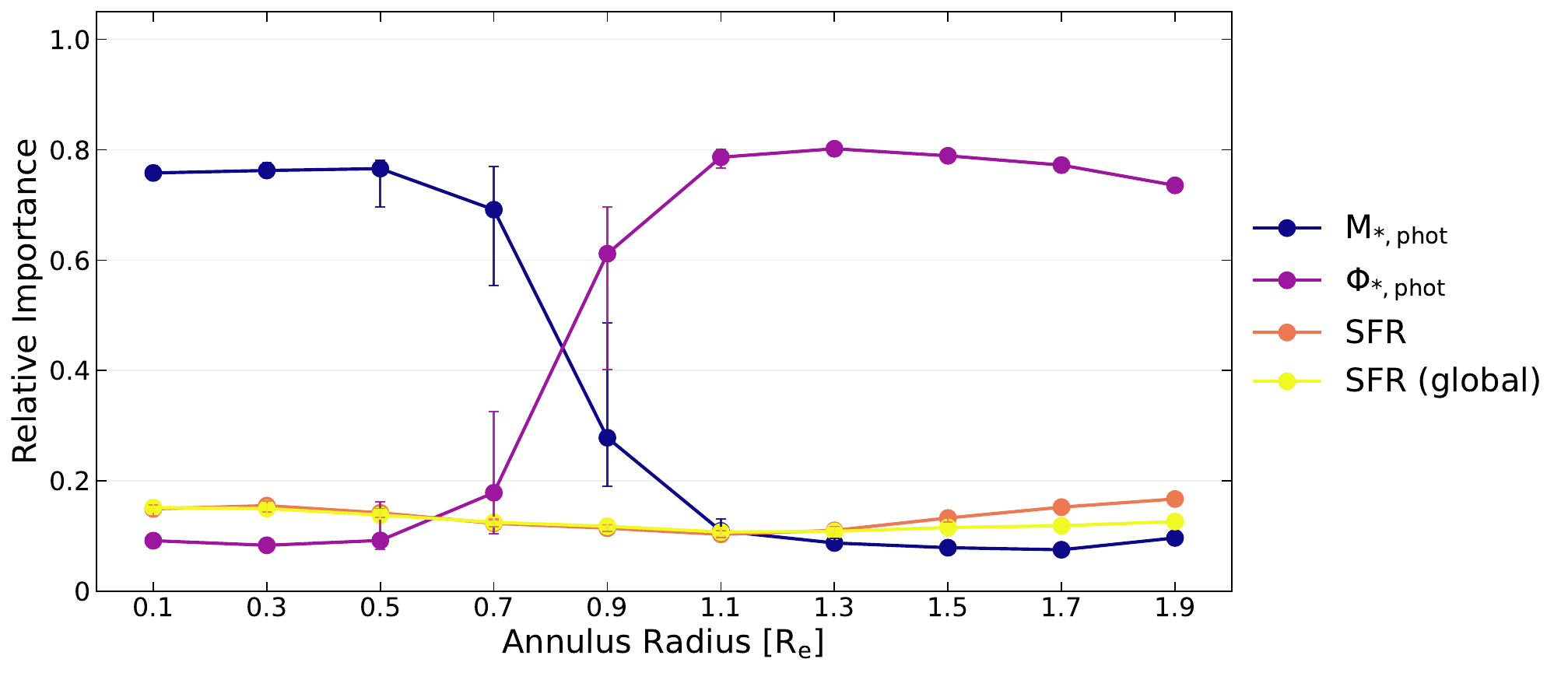}
    \caption{Importances in driving the gas-phase metallicity in annuli (of width $\rm 0.2 R_e$), as a function of Galactocentric radius of the annulus, of the photometric stellar mass $\mathrm{M_{*,phot}}$, photometric baryonic gravitational potential $\mathrm{\Phi_{*,phot}}$, global $\mathrm{SFR}$ from all star-forming spaxels from the pyPipe3D catalogue \citep{sanchez_sdss-iv_2022}, and integrated $\mathrm{SFR}$ obtained by taking the sum of selected star-forming spaxels within the same annuli. 
    The x-axis corresponds to the radial position of the centre of each $\rm 0.2 R_e$ annulus. The results indicate a spatial variation: at smaller radii, the stellar mass dominates over the baryonic gravitational potential, but this flips toward larger radii ($\mathrm{\gtrsim 0.9 R_e}$).}
    \label{fig:rf_manga_radii}
\end{figure*}

The results on the relative importance of these parameters in determining the metallicity within $\rm 1R_e$ are shown in Figure~\ref{fig:manga_rf}, in the form of a bar chart. The stellar mass estimated via spectro-photometric measurements is the main driver of the metallicity. Stellar mass being the main driver of the gas-phase metallicity was also confirmed by \citet{baker_stellar_2023}, although for the spectroscopic measurement of the stellar mass (their analysis did not include the photometric stellar mass estimates) and for a dataset that is only about a quarter the size of ours. 
There is a secondary dependence of $\mathrm{\sim 17 \%}$ on the stellar gravitational potential inferred via the spectro-photometric stellar mass, but a significantly larger error bar also accompanies it. Stellar mass and gravitational potentials inferred from pure spectroscopy are ruled out, and the dynamical properties (within 1R$_e$) also have insignificant relative importance, eliminating them as possible drivers for the gas-phase metallicity within $\rm 1R_e$. 

Spectro-photometric estimates of stellar mass and gravitational potential overpowering estimates from IFU spectroscopy were also confirmed by \citet{sanchez-menguiano_stellar_2024}. This likely reflects the fact that photometric masses, derived from broadband SED fitting, typically achieve higher S/N than spectroscopic masses, which rely on narrower spectral features that can be more sensitive to noise and template mismatches \citep{kauffmann_host_2003, drory_comparing_2004}. In addition, while photometric masses may be less accurate in an absolute sense due to modelling assumptions and systematic biases, they tend to be more precise, yielding consistent results across the galaxy population \citep{sanchez-menguiano_stellar_2024}. Tree-based algorithms like random forests naturally favour features with lower noise and higher variance, which may explain why photometric masses are given greater importance in our analysis. However, in contrast to our results, \citet{sanchez-menguiano_stellar_2024} find that $\mathrm{\Phi_{,phot}}$ ranks significantly higher than $\mathrm{M_{,phot}}$.

Other previous studies also found a strong dependence on $\mathrm{\Phi_*}$  \citep{deugenio_gas-phase_2018, cappellari_full_2023,sanchez-menguiano_stellar_2024}. Generally, this importance is explained by $\mathrm{\Phi_*}$ being a tracer of the gravitational potential of the galaxy and being connected to retaining metals. However, we also include $\mathrm{\Phi_{dyn,Re}=M_{dyn}/R_e}$ in our analysis, which should give a more accurate measure of the gravitational potential than $\mathrm{\Phi_*}$. Yet, $\mathrm{\Phi_{dyn,Re}}$ has a negligible low relative importance according to our random forest analysis shown in Figure~\ref{fig:manga_rf}. We further investigate the correlation between $\mathrm{M_{dyn,R_e}}$ and the gas-phase metallicity via a differential uncertainty measurement test in Appendix~\ref{sec:app_diff_unc}.

Part of the discrepancies between previous studies on the relative importance of stellar mass versus gravitational potential were possibly due to the use of different methodologies to measure the stellar mass, as discussed above. However, in the next section, we discuss the more prominent role of spatial scales as the origin of the apparent discrepancies.

We conclude this subsection by noting that the third-ranked property in predicting the metallicity is the SFR. This is not surprising, as it was expected from the FMR. However, we will discuss this aspect later on in a separate section.

\subsubsection{Radial dependence: $\rm M_*$ overthrown by $\Phi_*$ at large radii}\label{sec:results_radial}

Prompted by the puzzling discrepancies between different studies on the relative importance of stellar mass and gravitational potential in determining the gas metallicity, we have explored the role of the spatial scale on which such importance is assessed. More specifically, we extended the analysis performed in the previous section, which was on the metallicities estimated within $\rm 1R_e$, by re-running our random forest regression for average metallicities within narrow annuli of width $\rm 0.2 R_e$ from the centre of each galaxy out to $\rm 2R_e$.


Since we have already established that the stellar mass and gravitational potential from photometry, and the SFR are the three most important drivers of the gas-phase metallicity, we solely focus on these parameters. The results are shown in Figure~\ref{fig:rf_manga_radii}. It is evident that both $\mathrm{M_*}$ and $\Phi_*$ have a radial dependence. At small radii, the stellar mass is by far the dominant quantity determining the metallicity. As the radius increases, the metallicity is less impacted by changes in the stellar mass but increasingly correlates with the gravitational potential. We also include SFRs computed within the annuli and the global SFR taken from the pyPipe3D catalogue in our analysis, but find no significant difference in the role of SFR as a function of radius. 

The most likely interpretation is that in the inner regions of galaxies, the strong correlation between metallicity and stellar mass arises because stellar mass traces the integrated metal production from the star formation history \citep{sanchez_characteristic_2014, peng_strangulation_2015, chisholm_metal-enriched_2018, maiolino_re_2019}. Central regions, characterised by deep gravitational potentials and high stellar densities, have short chemical enrichment timescales and efficient recycling of metals through stellar feedback, enabling evolution with little metal loss regardless of the overall gravitational potential \citep{baker_stellar_2023}.

In the outer regions, where metals are less gravitationally bound, the gravitational potential (of which $\Phi_*$ is a proxy) becomes more important in regulating gas-phase metallicity. These regions are more susceptible to gas inflows, feedback-driven outflows, and radial migration, all of which can dilute or redistribute metals \citep{fraternali_gas_2017, christensen_-n-out_2016, tumlinson_circumgalactic_2017}. The longer chemical mixing and cycling timescales in the outskirts \citep{ho_metallicity_2015, tissera_oxygen_2019} mean they may not reach chemical equilibrium, making gravitational potential a better predictor of local metallicity than stellar mass.

We finally note that the importance of dynamical gravitational potential ($\rm \Phi _{dyn,R_e}$) is minimal, and remains so regardless of the radius within which the metallicity is estimated. This is likely because the gravitational potential was estimated by \citet{zhu_manga_2023} within $\rm 1R_e$, in the region where stellar mass dominates the importance. Unfortunately, dynamical masses estimated at larger radii are not available.

\subsubsection{The relative importance in integrated SDSS spectra}

\begin{figure}
    \centering
    \includegraphics[width=\linewidth]{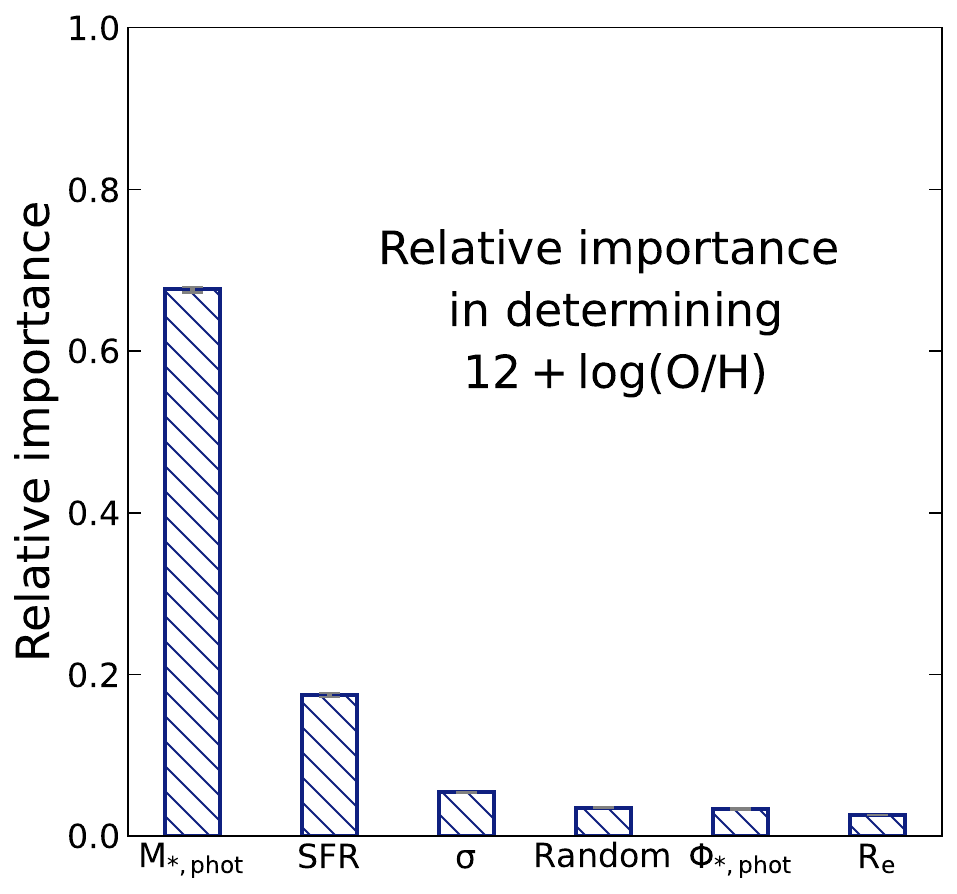}
    \caption{Results for the random forest regression for our SDSS sample. As there are no dynamical or spectroscopic mass measurements available, we only include the following parameters: photometric stellar mass $\mathrm{M_{*,phot}}$, $\mathrm{SFR}$, velocity dispersion $\mathrm{\sigma}$, baryonic gravitational potential $\mathrm{\Phi_{*,phot}}$, effective radius $\mathrm{R_e}$, and a random uniform control variable. The error bars represent the 16th and 84th percentiles from 100 bootstrap random samples. Our results again indicate that the gas-phase metallicity is most dependent on the stellar mass.}
    \label{fig:rf_sdss}
\end{figure}

We complement our analysis of 14 different parameters for the MaNGA survey by conducting the same random forest for $\sim 120,000$ galaxies from SDSS.
SDSS spectra offer far larger statistics, but provide only information integrated within the fibre's aperture.
The results are shown in Figure~\ref{fig:rf_sdss} and confirm our previous finding of the photometric stellar mass $\mathrm{M_{*,phot}}$ being the major driver of the gas-phase metallicity, while $\mathrm{\Phi_{*,phot}}$ plays a minor role. This is essentially the same result as obtained in Manga when restricting the metallicity measurements to within $\rm 0.5R_e$. We do not have spatially resolved information in this case; however, we have tested the random forest by selecting samples for which the fibre's aperture includes gradually a larger fraction of the galaxy, out to $\rm 2R_e$ and found no significant variation in the importance of $\Phi_*$. The reason is likely that, regardless of the fraction of the galaxy enclosed in the aperture, the nebular emission lines used for determining the metallicity are dominated by the inner (brighter) region, where there is the highest surface density of star formation. As a consequence, the inferred metallicities are typically dominated by the central regions, hence explaining the dominance of $\rm M_*$ according to the MaNGA result in Fig.\ref{fig:rf_manga_radii}.



\subsubsection{The role of SFR}\label{sec:results_SFR}

As previously mentioned, we do find that the SFR plays a role in determining metallicity, consistent with expectations from the FMR. This role does not vary significantly as a function of the radius within which the metallicity is estimated. If the dependence on SFR is driven by gas accretion, which both dilutes the metallicity and boosts the star formation, then this would imply that this process is ongoing on all galactic scales. This would be in line with independent, detailed studies of extraplanar gas kinematics in some nearby galaxies \citep[e.g.][]{li_kinematic_2021,li_fountain-driven_2023}.

It is, however, somewhat puzzling that the importance of the SFR is significantly lower than the $\sim 30\%$ originally estimated by the FMR \citep{mannucci_fundamental_2010,curti_mass-metallicity_2020}. Using MaNGA metallicities within $\rm 1R_e$, the metallicity importance is only 7\% (Fig.\ref{fig:manga_rf}). At $\rm 0.5R_e$ the importance of SFR increases slightly, but it is still only about 15\% (Fig.\ref{fig:rf_manga_radii}). Note that this is regardless of whether we use the total SFR in the galaxy (yellow curve in Fig.\ref{fig:rf_manga_radii}), or the SFR estimated within a given radius (orange curve in Fig.\ref{fig:rf_manga_radii}). Even in the full SDSS sample, the role of the SFR is still below 20\% (Fig.\ref{fig:rf_sdss}). A likely explanation is that in the original FMR works, where only $\mathrm{M_*}$ and SFR were considered, the SFR indirectly picked the metallicity dependence on other galaxy parameters that were not included in the original analysis, such as $\Phi$ and $\sigma$, which are instead considered in these RF studies. This is also supported by the fact that the importance of $\mathrm{M_*}$, at the level of $\sim$70\%, is instead consistent with the original FMR studies.

However, the scenario is probably even more complex, as it emerges from the analysis of the Partial Correlations, as discussed in the next section.

\subsection{Partial Correlation Coefficients: Sign of Correlations}

In this section, we explore the signs of the dependence of the metallicity on the galactic parameters by leveraging partial correlations. Partial correlations may also help visualise better the causal relations with the aid of 3D plots, i.e. by illustrating the distribution of galaxies as a function of two quantities, colour-coded by metallicity. However, one should be aware that this approach is limited to the comparison between two specific quantities, ignoring others that might also be relevant.

Given that we have already assessed the dominance of the spectro-photometric masses relative to those derived via IFS-spectroscopy, we do not consider the latter in this analysis. Additionally, since $\rm \Phi_* = M_*/R_e$ contains the stellar mass, we cannot use these two quantities together in the analysis of the partial correlation coefficients. However, in the next section, we illustrate methodologies for still investigating the dependence of the metallicity sign on $\mathrm{\Phi_*}$.

Additionally, we note that a difference in correlation strengths obtained by random forest regression PCCs is expected: PCCs cannot manage non-monotonic relations, but random forests can. Furthermore, random forests are better at disentangling intrinsic drivers. Therefore, discrepancies might indicate some non-monotonic correlations among our parameters. Nonetheless, as we will see, both methods agree that the stellar mass is the key driver of the gas-phase metallicity. We rely on the random forest to reliably measure the strength of the relations, while the PCCs tell us the signs of correlations. 

\begin{figure*}
    \centering
    \includegraphics[width=0.49\linewidth]{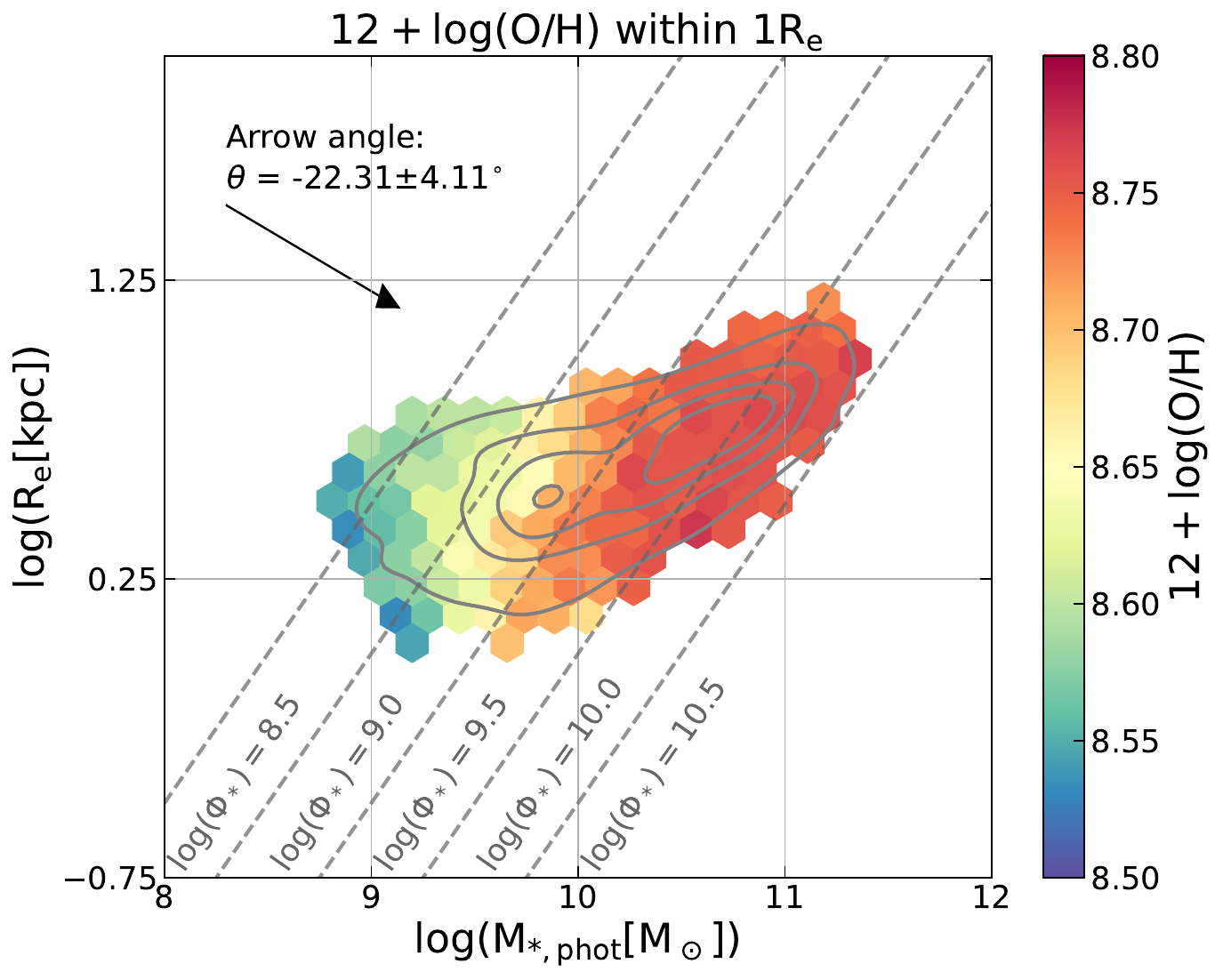}
    \includegraphics[width=0.49\linewidth]{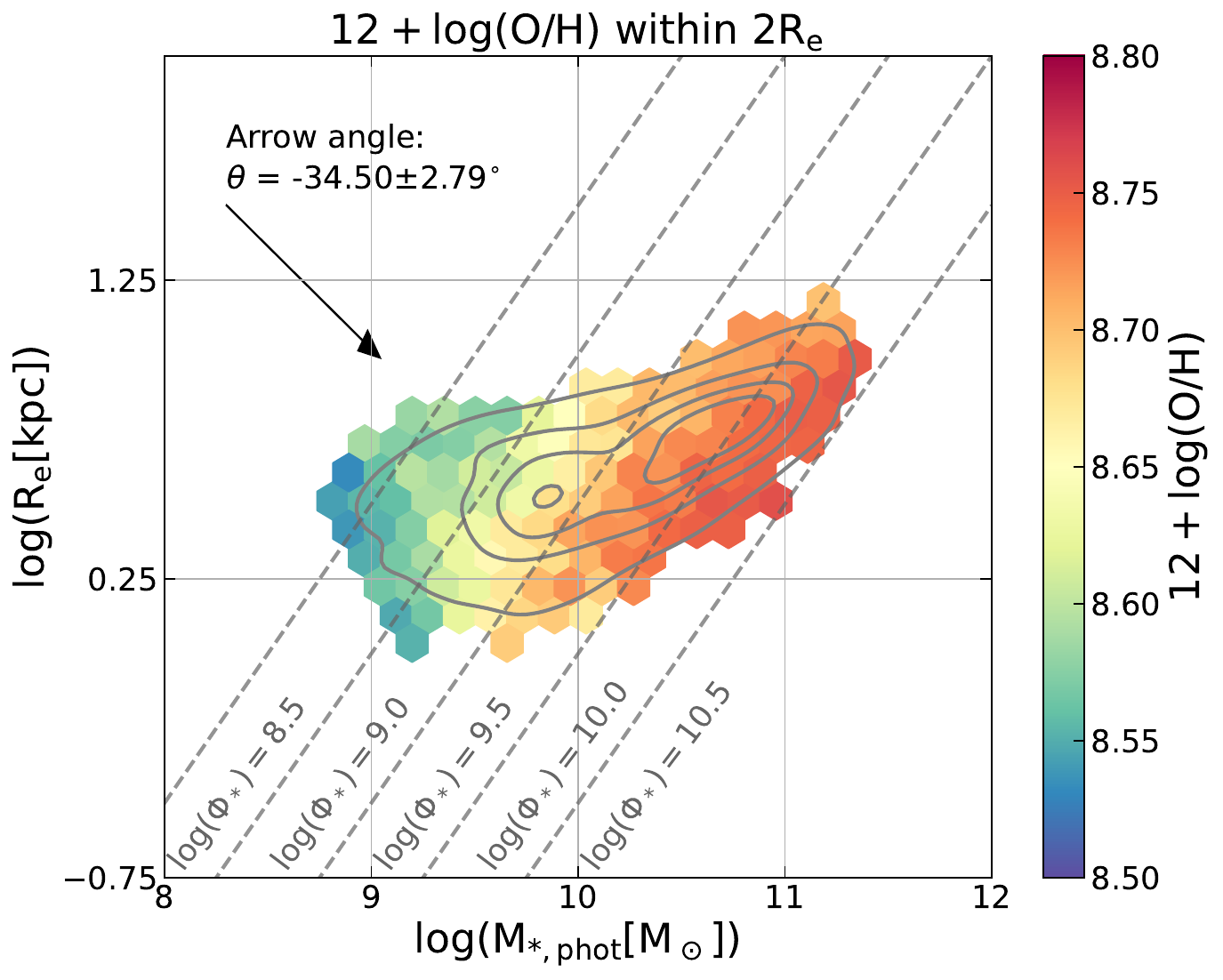}
    \caption{2D histogram of the effective radius versus photometric stellar mass, colour-coded by the gas-phase metallicity for the MaNGA survey. The data is plotted in hexagonal bins, and the grey density contours highlight the distribution of galaxies, with the outer contour corresponding to $90 \%$ the data. The colour coding of each bin gives the average metallicity of the galaxies in that bin; left is for the metallicity within $\rm 1R_e$, while right is for the metallicity within $\rm 2R_e$. The dashed lines indicate the locii at constant $\rm \Phi_*$. The arrow angles are defined in Eq.\ref{eq:arrow_angle} and indicate the direction of the average gradient.}
    \label{fig:hexbin_Re_Mst}
\end{figure*}

\subsubsection{The dependence on the stellar gravitational potential}

Fig.\ref{fig:hexbin_Re_Mst} shows the distribution of MaNGA galaxies on the $\rm R_e$ vs $\rm M_*$ plane, colour-coded by metallicity within $\rm 1R_e$ (left) and within $\rm 2R_e$ (right). The data are binned in $\rm R_e$ and $\rm M_*$, such that each bin includes at least ten galaxies, and the colour coding gives the average metallicity in each bin. The contours show the distribution of galaxies, where the outermost contour contains 90\% of the galaxies.
Exploring the average metallicities in bins rather than the distribution of individual galaxies avoids the partial correlation analysis being dominated by the region of the plane that is most populated. 

Clearly, the gradient in the diagram is dominated by $\mathrm{M_*}$, not unexpectedly given the MZR and the results from the random forest. However, the colour shading is tilted, indicating a secondary, inverse dependence on $\mathrm{R_e}$, which is tracing the dependence on $\mathrm{\Phi_*}$ identified in the random forest. Indeed, the dashed lines show the locii of constant $\mathrm{\Phi_*}$, and these are nearly parallel to the iso-metallicity profiles. This is quantified by the arrow angles, derived from Eq. \ref{eq:arrow_angle}, which essentially give the direction of the average gradient of the metallicity on the $\mathrm{M_*-R_e}$ plane. An arrow angle equal to zero (horizontal gradient) would imply a complete relative dependence only on $\rm M_*$ and no dependence on $\rm \Phi_*$. An arrow angle of $-45^\circ$ would indicate total dependence on $\mathrm{\Phi_*}$. In Fig.\ref{fig:hexbin_Re_Mst} we see that for the metallicities within $\rm 1R_e$ the arrow angle is $\sim -23^\circ$, meaning a strong, positive dependence on stellar mass and some additional positive dependence on $\mathrm{\Phi_*}$. When considering the metallicities within $\rm 2R_e$, the arrow angle becomes more negative ($-35^\circ$) and approaches $-45^\circ$, indicating that the positive dependence on $\mathrm{\Phi_*}$ becomes stronger. These findings are fully consistent with what was found in the random forest analysis, but now with the signs of the correlations.

\begin{figure*}
    \centering
    \includegraphics[width=0.49\linewidth]{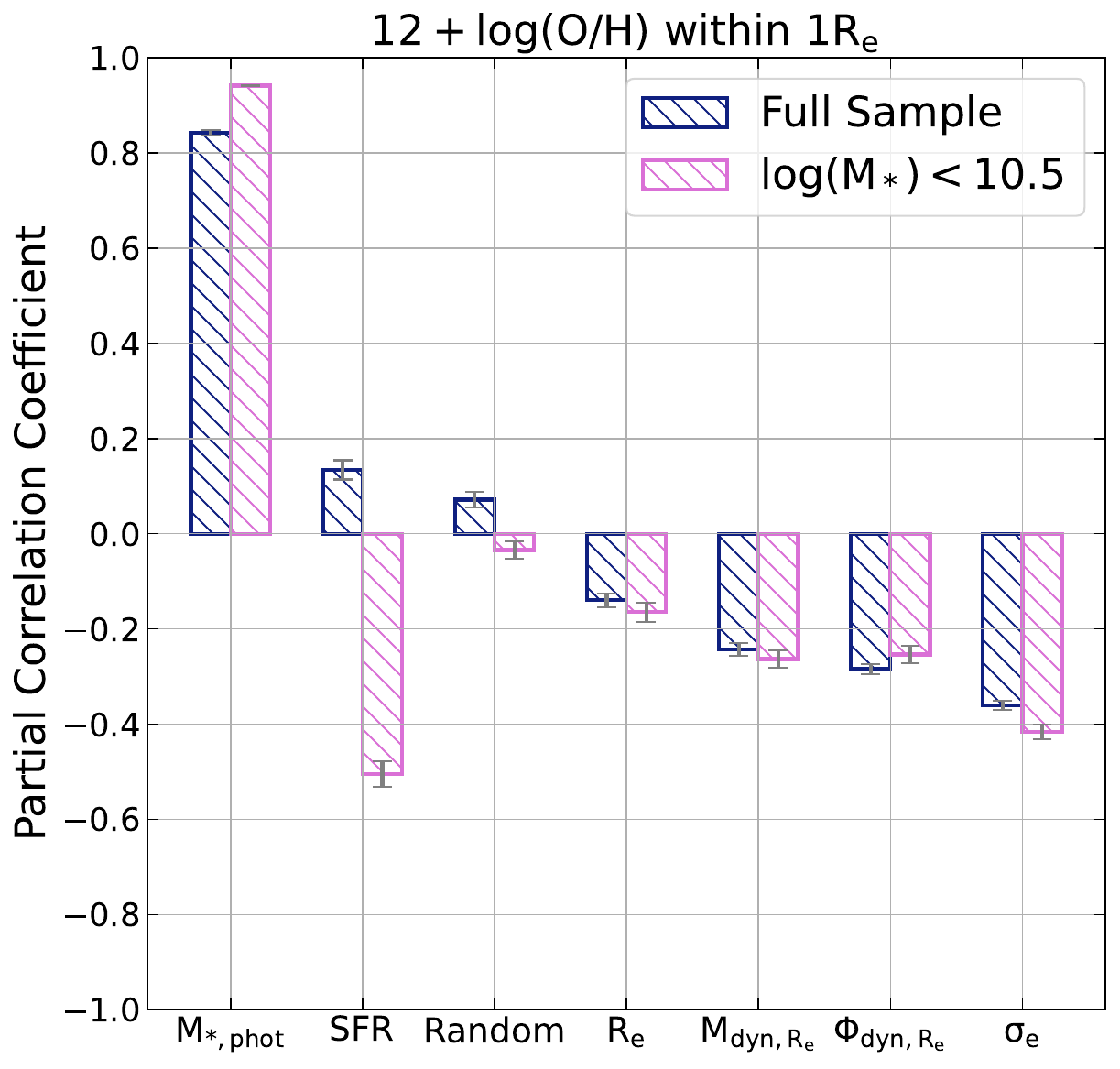}
    \includegraphics[width=0.49\linewidth]{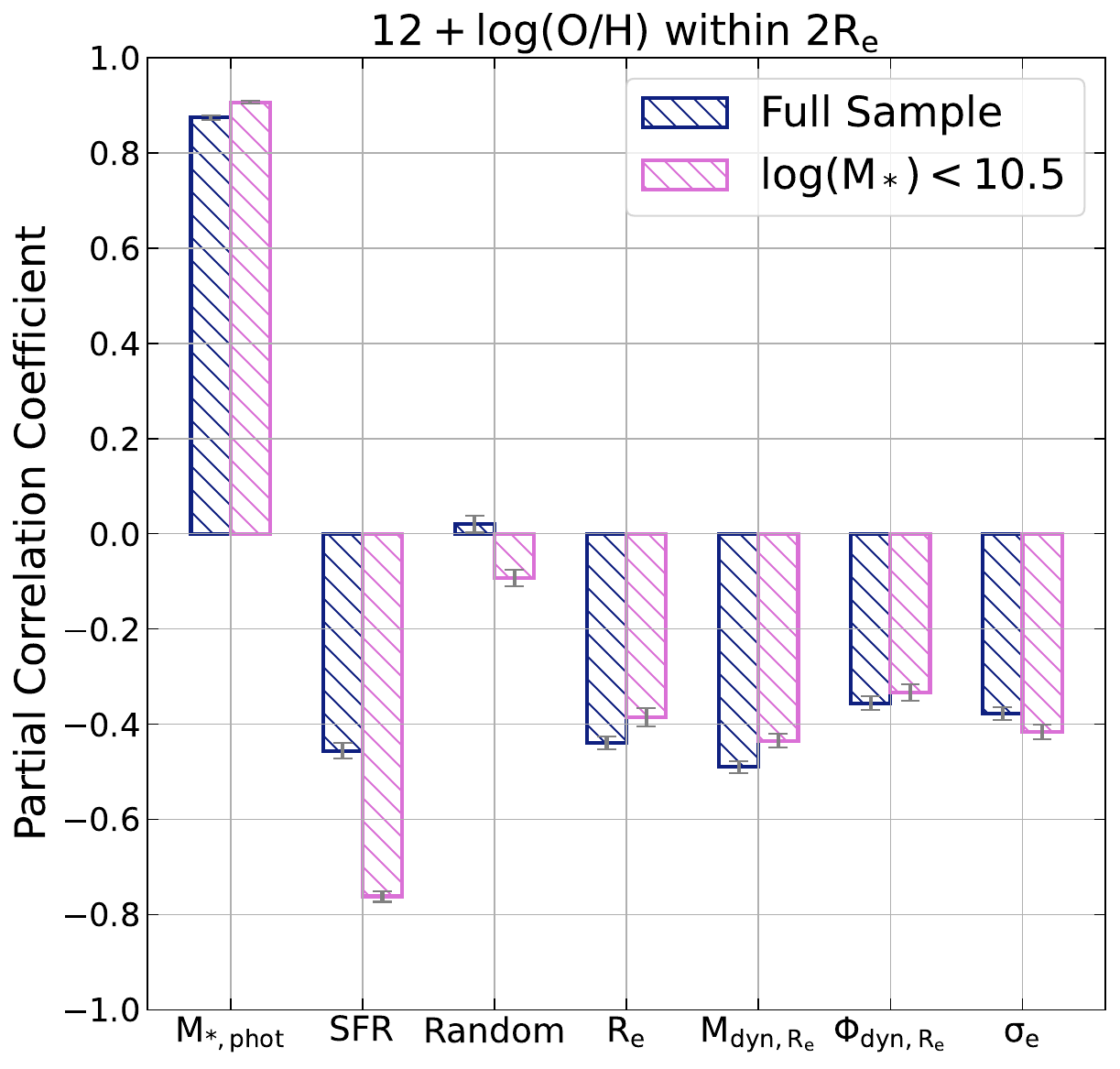}
    \caption{Partial Correlation Coefficients (PCC) between the gas-phase metallicity and several galaxy properties from the MaNGA survey: photometric stellar mass $\mathrm{M_{*,phot}}$, SFR, effective radius $\mathrm{R_e}$, dynamical mass $\mathrm{M_{dyn}}$, dynamical gravitational potential $\mathrm{\Phi_{dyn}}$, velocity dispersion $\mathrm{\sigma_e}$, and a uniform random variable (Random). Left is for the metallicities within $\rm 1R_e$, and right is for the metallicities within $\rm 2R_e$. Blue histograms are for the full sample, while violet histograms are limited to galaxies within $\mathrm{\log{(M_*/M_\odot)}<10.5}$.}
    \label{fig:pcc_manga}
\end{figure*}

\begin{figure*}
    \centering
    \includegraphics[width=0.49\linewidth]{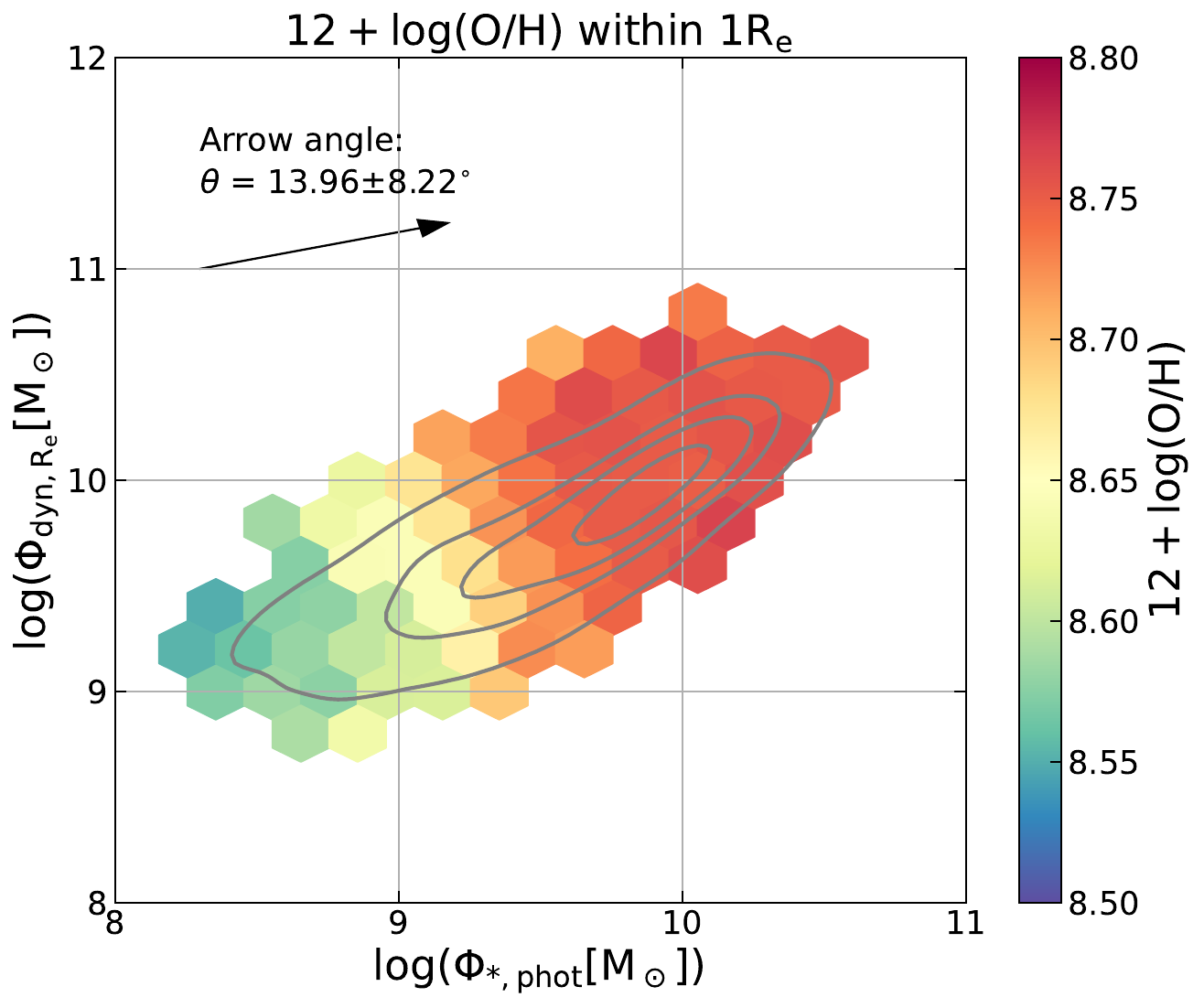}
    \includegraphics[width=0.49\linewidth]{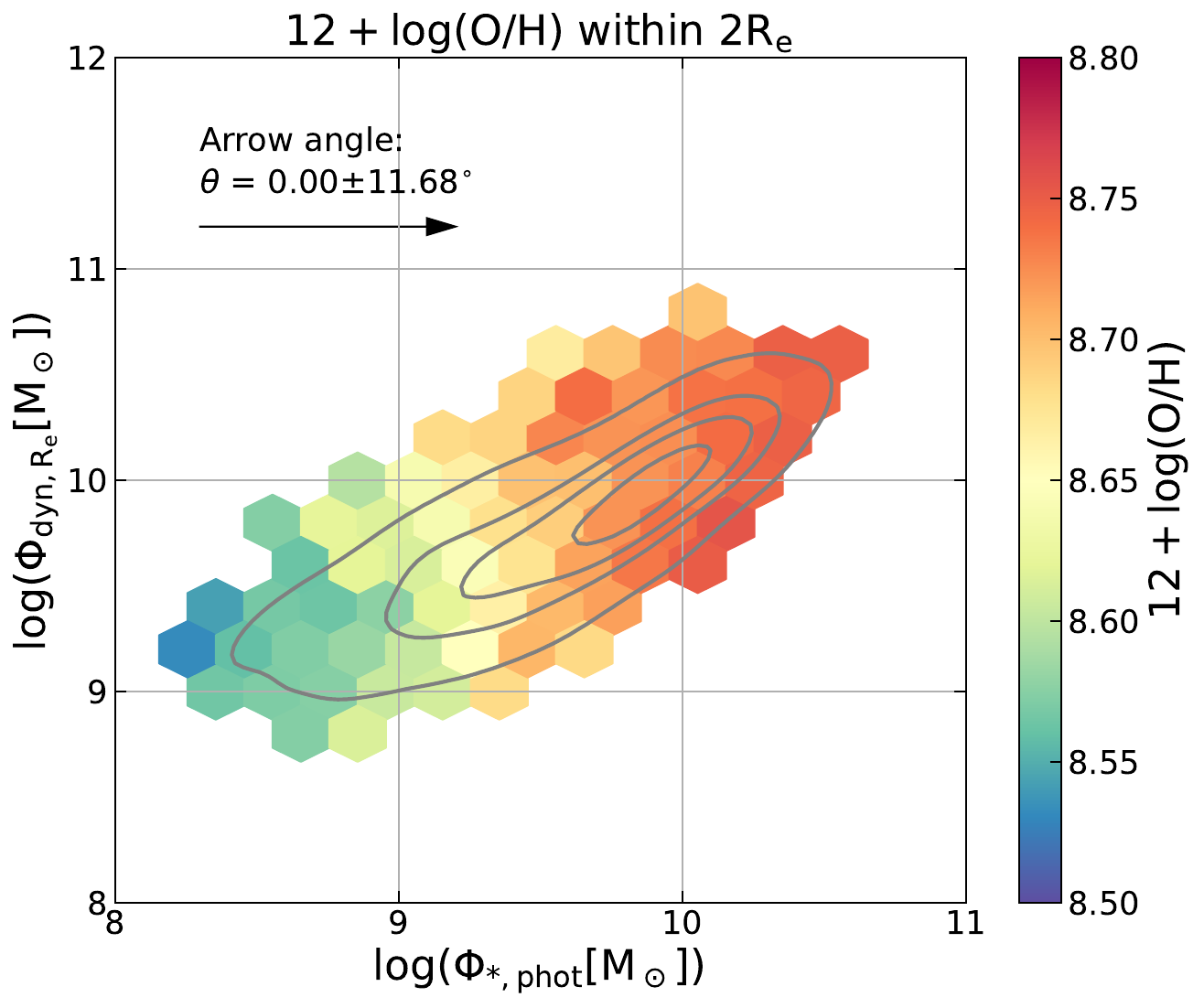}
    \caption{
    As Fig.\ref{fig:hexbin_Re_Mst} but where we now show the MaNGA galaxies in the $\rm \Phi_{dyn,R_e}$ versus $\rm \Phi_{*,phot}$ plane, colour coded by metallicity. Left is for the metallicity within $\rm 1R_e$ while right is for the metallicity within $\rm 2R_e$.}
    \label{fig:hexbin_Phi_Phi}
\end{figure*}

Fig. \ref{fig:pcc_manga} illustrates the results, including the signs, of the PCCs between metallicities and various galactic quantities in the MaNGA sample. As usual, the metallicity shows the strongest (positive) correlation with stellar mass.
As discussed, $\mathrm{\Phi_*}$ cannot be included here, because it contains $\rm M_*$. However, here we are showing $\mathrm{M_{dyn,R_e}}$ and $\mathrm{\Phi_{dyn,R_e}}$, which we recall are calculated within $\rm 1R_e$. The metallicity shows a secondary dependence on these quantities; however, on the contrary to what was expected from the very positive dependence on $\mathrm{M_*}$ and $\mathrm{\Phi_*}$ discussed above (and shown in Fig.\ref{fig:hexbin_Re_Mst}), the dependence on $\mathrm{M_{dyn,R_e}}$ and $\mathrm{\Phi_{dyn,R_e}}$ are negative. We explore this effect more closely in Fig.\ref{fig:hexbin_Phi_Phi}, where we show the 2D plot of the distribution of galaxies binned in $\mathrm{\Phi_{dyn,R_e}}$ and $\mathrm{\Phi_*}$, colour coded by metallicity within $\rm 1R_e$ and $2\rm R_e$. This diagram clearly illustrates the much stronger positive dependence on $\mathrm{\Phi_*}$ relative to $\mathrm{\Phi_{dyn},R_e}$ in both cases. When controlling for $\mathrm{\Phi_*}$, the dependence on $\mathrm{\Phi_{dyn,R_e}}$ for the metallicity within $\rm 1R_e$ is positive, while null for the metallicity within $\rm 2R_e$. This striking difference between the behaviour relative to $\mathrm{\Phi_{dyn,R_e}}$ and $\mathrm{\Phi_*}$ is puzzling, as they should both be tracing the gravitational potential. While in the representation of Fig.\ref{fig:hexbin_Phi_Phi} $\mathrm{\Phi_*}$ may have a strong contribution from the positive dependence on $\mathrm{M_*}$, that cannot be the entire explanation, given that we have seen in Fig.\ref{fig:hexbin_Re_Mst} and in the previous section, that there is also a strong direct and casual dependence on $\mathrm{\Phi _*}$. Additionally, we have seen in Fig.\ref{fig:pcc_manga} that there is a {\it negative} metallicity correlation also with $\mathrm{M_{dyn,R_e}}$, regardless of $\mathrm{\Phi_{dyn,R_e}}$. The key to understanding the puzzle probably is in the fact that $\mathrm{\Phi_{dyn,R_e}}$ is determined within $\rm 1R_e$, while $\mathrm{\Phi_*}$ is determined by taking the whole mass of the galaxy, even well outside $\rm 2R_e$. Therefore, $\mathrm{\Phi_*}$ is probably a proxy of the global gravitational potential of the galaxy, while $\mathrm{\Phi_{dyn,R_e}}$ is only tracing the gravitational potential within the inner ($\rm 1R_e$) region. Therefore, the strongly positive metallicity correlation with $\mathrm{\Phi_*}$ may indeed reflect the global capability of the galaxy in retaining metals. The inverse dependence on $\mathrm{\Phi_{dyn,R_e}}$ is more difficult to understand, but has also been found by a much smaller sample in \citet{baker_stellar_2023} (see Fig. 2 of that work). We speculate that a deeper inner gravitational potential may be the sink for a stronger inflow of metal-poor, or nearly pristine, gas towards the central region of the galaxy, which would dilute the metallicity, hence resulting in a negative correlation. This would be a local, more subtle version of a similar effect seen more prominently at high-z, in the form of inverted metallicity gradients \citep{cresci_gas_2010, troncoso_metallicity_2014}.

\subsubsection{The dependence on SFR}

\begin{figure}
    \centering
    \includegraphics[width=\linewidth]{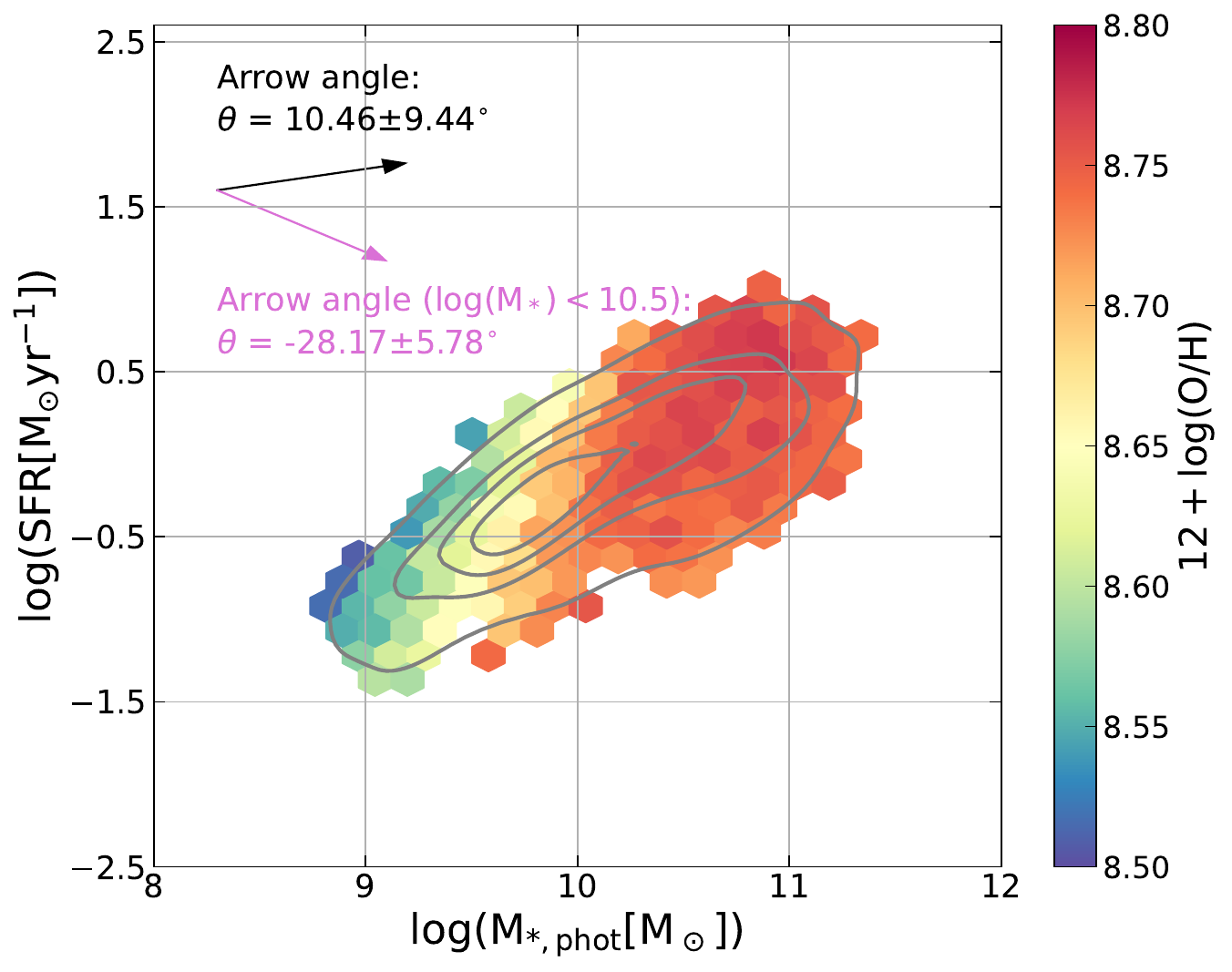}
    \caption{2D histogram of the photometric stellar mass versus SFR colour-coded by the gas-phase metallicity for the MaNGA survey. The data is plotted in hexagonal bins, and the grey density contours highlight the distribution of galaxies, with the outer contour corresponding to $90 \%$ the data. We include a PCC arrow pointing into the strongest increasing gradient in gas-phase metallicity. Two arrows are included: for the whole selected MaNGA sample (black) and for adding an additional cut of $\mathrm{\log(M_{*,phot})<10.5}$. For the latter arrow, we see indications of the FMR: for a given stellar mass, the SFR and metallicity are anti-correlated.}
    \label{fig:hexbin_manga}
\end{figure}

We finally discuss the dependence on the SFR. In section~\ref{sec:results_SFR}, the RF confirmed a dependence on SFR, which is actually of third importance after accounting for $\mathrm{M_*}$ and $\Phi_*$. Here we explore the sign of the dependence by leveraging the PCC information. Fig.\ref{fig:pcc_manga}-right shows a strong negative metallicity dependence (i.e. anti-correlation) with SFR, as expected from the FMR. However, this is true only for the metallicity estimated within $\rm 2R_e$ (right panel of Fig.\ref{fig:pcc_manga}). When considering the metallicity within $\rm 1R_e$ (left panel of Fig.\ref{fig:pcc_manga}), the sign of the correlation for the full sample becomes positive, i.e. opposite of what is expected from the FMR. We explore this phenomenon further in the 2D plot of Fig.\ref{fig:hexbin_manga}, which shows the distribution of MaNGA galaxies in the SFR versus $\rm M_*$ diagram, colour coded by metallicity within $\rm 1R_e$. The inverted dependence is driven by the very massive galaxies, which is where the mass-metallicity relation saturates, or becomes even slightly inverted, as already pointed out in the past by other authors \citep{yates_relation_2012, peng_dependence_2014, duarte_puertas_mass-metallicity_2022}; this is because the metallicity reaches the effective yield, beyond which additional star formation or mass growth does not significantly raise the metallicity \citep{tremonti_origin_2004, peng_dependence_2014}. The possible inversion of the FMR has also been attributed to gradual dilution following gas-rich mergers \citep{yates_relation_2012}, and environmentally enriched inflows in satellites residing in dense regions \citep{peng_dependence_2014}. If we restrict the sample to galaxies less massive than $\mathrm{\log{(M_*/M_\odot)}<10.5}$, then the dependence on the SFR is back to the negative value expected from the FMR; this is illustrated by the magenta arrow angle in Fig.\ref{fig:hexbin_manga} and by the magenta histogram in Fig.\ref{fig:pcc_manga}. Notably, excluding high mass galaxies from our sample also results in significantly lower errors for the arrow angle. 

It is, however, puzzling that the random forest analysis of the full sample did not show any variation of the dependence on the SFR as a function of radius (Fig.\ref{fig:rf_manga_radii}). The reason is probably that the random forest result is dominated by the bulk of the galaxies being at $\mathrm{\log{(M_*/M_\odot)}<10.5}$, as illustrated by the contours in Fig.\ref{fig:hexbin_manga}, where the dependence on the radius is much reduced, as seen by the magenta histograms in Fig.\ref{fig:pcc_manga}.

\section{Discussion}\label{sec:discussion}

Our results have shown that the apparent discrepancies between previous studies on the primary drivers of the gas-phase metallicity are primarily due to different definitions for the stellar mass, gravitational potential, and on the different spatial scales investigated. Once these aspects are taken into account, various previous studies are broadly consistent. To illustrate that there are no additional issues playing a role in the differences between previous studies, in Appendix~\ref{sec:app_reprod} we discuss that we can reproduce the results of previous works when their assumptions are adopted.

Having understood the relevance of different definitions of galactic quantities and having also assessed the key role of the spatial scales in understanding the parameters driving the metallicity, we can now use our results to obtain a more coherent picture of the physical processes driving the metallicity of galaxies.

One of the most important results is that stellar mass $\rm M_*$ and stellar gravitational potential $\rm \Phi_*$ (determined through the spectro-photometric method) are the most important parameters in determining the gas metallicity in galaxies. However, their relative importance swaps as a function of the radius within which the metallicity is determined. In the central region ($\rm <1~R_e$) stellar mass is the most important parameter determining the metallicity, while gravitational potential plays a minimal role. Our interpretation is that in the central region the potential well is always deep enough to retain metals, regardless of the large-scale gravitational potential. As a consequence, the amount of metals simply reflects the integrated production of stars, i.e. the stellar mass.
On large scales ($\rm > 1.5 R_e$), gravitational potential (as traced by $\rm \Phi_*$) is the most important parameter driving the metallicity. In such outer regions, the gravitational potential likely plays a prominent role in the retention of metals.

The role of the SFR (the third most important parameter in determining the metallicity) does not have a strong dependence on the spatial scale, especially when considering galaxies with masses smaller than $\rm 10^{10.5}M_\odot$. If the metallicity's inverse dependence on the SFR is interpreted as dilution from accreting pristine, or near-pristine gas, then our result may indicate that such accretion happens at the same level across the entire disc, as indeed observed in detailed studies of some individual nearby galaxies \citep{ho_metallicity_2015, kreckel_mapping_2019, kreckel_temperature-based_2025}. It is also a very interesting finding that, at any scale, even if it is the third most important parameter, the SFR contributes only at the 7--15\% level to determining the gas metallicity, which is much less than estimated by previous studies on the FMR \citep{mannucci_fundamental_2010, curti_mass-metallicity_2020}. The likely origin of this difference is that most of the previous studies had only considered stellar mass and SFR, and not the several other parameters considered here (gravitational potential, velocity dispersion, size, etc.); as a consequence, in those previous studies SFR may have picked the importance in determining the metallicity from the parameters that were not included in their analysis, and which probably correlate with the SFR. 

Finally, the finding that dynamical mass or total gravitational potential, as inferred from the dynamical mass, contribute very little to the metallicity of the galaxy, can be understood in terms of these quantities being derived only within $\rm 1R_e$. The fact that these dynamical tracers within $\rm 1R_e$ play little role in determining the metallicity further supports the scenario in which within the central region galaxies chemically evolve without loss of metals (or manage to reaccrete them locally), regardless of the depth of the gravitational potential in these regions.

Overall, our study highlights the importance of assessing the metallicity driving factor on different galactic scales. Determining the role of different galactic properties in driving the gas metallicity without taking into account the spatial component of these relations may lead to a partial or even incorrect view of the associated physical processes. Our study has also revealed that the accuracy with which the galactic properties are measured (in particular, the stellar mass) has an impact on the capability of disentangling causation from correlation.

\section{Conclusions}\label{sec:conclusion}

We have investigated the correlations between the gas-phase metallicities and other galaxy parameters such as the stellar mass $\mathrm{M_*}$, baryonic gravitational potential $\mathrm{\Phi_*}$, dynamical mass $\mathrm{M_{dyn}}$, and $\mathrm{SFR}$ (and many others) using large samples from the MaNGA and SDSS surveys. We have applied random forest regression and Partial Correlation Coefficients to unravel the relative importance of different parameters in determining the gas-phase metallicity, as well as the signs of the correlations. 

We have found that apparent discrepancies between previous studies are mostly due to different ways of defining the galactic properties (and in particular the stellar masses and gravitational potentials) and to the different spatial scales that were investigated. 
Once these aspects are properly taken into account, we find the following key results and corresponding possible scientific conclusions:

\begin{itemize}
    
    \item We find that the primary drivers of the metallicity are stellar mass, $\mathrm{M_*}$, and the stellar gravitational potential, $\mathrm{\Phi_* = M_*/R_e}$ (which is likely a proxy of the total gravitational potential). However, we find that their relative role has a strong radial dependence: at small galactocentric distances ($\mathrm{\lesssim 0.7 R_e}$), metallicity is primarily driven by stellar mass, whereas at larger radii ($\mathrm{\gtrsim 0.9 R_e}$), $\mathrm{\Phi_*}$ becomes the dominant factor. We attribute this to the fact that central regions of galaxies experience rapid metal enrichment and efficient retention due to short chemical timescales and deeper gravitational wells, while in the outskirts of galaxies, chemical mixing is less efficient and gravitational potential is more influential in regulating outflows (hence metal retention) and, possibly, also inflows (hence metals dilution). 

    \item The third most important parameter (after $\rm M_*$ and $\Phi_*$) in determining the metallicity is the SFR. We find that the SFR only accounts for 7--15\% in driving the metallicity; we argue that previous studies had found a more prominent role ($\sim 30\%$) because other galactic parameters were not included in the analysis. Additionally, the dependence on the SFR is complex and mass-dependent: while a clear negative correlation emerges when measuring metallicity within $\mathrm{2R_e}$, consistent with the FMR, the relation reverses within $\mathrm{1R_e}$. This inverted trend is likely driven by galaxies above $\mathrm{\log(M_*/M_\odot) \gtrsim 10.5}$, where metallicity approaches the effective yield limit. When limiting the sample to lower-mass galaxies ($\mathrm{\log(M_*/M_\odot) < 10.5}$), the expected FMR-like anti-correlation is recovered. The random forest did not reveal a radius-dependent SFR trend. This is likely explained by the dominance of lower-mass galaxies in the sample, which dilutes the radial variation in the global analysis.
    
    \item While there is a strong dependence of the metallicity on $\Phi_*$, especially at large radii,
    the gravitational potential measured within $1\mathrm{R_e}$ via the dynamical mass, $\mathrm{\Phi_{dyn,R_e}}$, shows a weak or even negative correlation with metallicity, which we speculate may be due to localised inflows of metal-poor gas into deeper inner potential wells. This suggests that while $\mathrm{\Phi_*}$ captures global retention effects, $\mathrm{\Phi_{dyn,R_e}}$ may instead trace local dilution processes, similar to inverted metallicity gradients observed at high redshift.

    \item Finally, we have found that the precision with which the quantities are measured also plays an important role. Specifically, the stellar mass (or baryonic potential) measured via photometry or spectro-photometry is the one that results in the strongest power in determining the metallicity, while spectroscopic measurements of the stellar masses have much smaller predictive power. This is likely due to the fact that spectroscopic stellar mass determinations are only confined to the central regions (either because the S/N is lower in the outer parts of the MaNGA spectra or, in the case of SDSS spectra, the single fibre only samples the central region), while photometric or spectro-photometric measurements can recover the stellar mass over the entire galaxy.
    
\end{itemize}

Our results provide an important key reference for comparison with models and simulations, and also reconcile tensions between previous studies. Our findings also provide a local benchmark for similar studies at high redshift.

\section*{Acknowledgements}
We are grateful to the referee, Jianhui Lian, for helpful comments that have improved the manuscript. We thank Mirko Curti for kindly providing the code used in our gas-phase metallicity measurements. We are very grateful for feedback and comments from Francesco Belfiore and Asa Bluck. MK thanks the University of Cambridge Harding Distinguished Postgraduate Scholars Programme, UK Science and Technology Facilities Council (STFC) Center for Doctoral Training (CDT) in Data Intensive Science, and Girton College Cambridge for a PhD studentship. RM acknowledges support from the Science and Technology Facilities Council (STFC), by the European Research Council (ERC) through Advanced Grant 695671 ``QUENCH'', by the UK Research and Innovation (UKRI) Frontier Research grant RISEandFALL. RM also acknowledges support from a Royal Society Research Professorship grant. WMB gratefully acknowledges support from DARK via the DARK fellowship. This work was supported by a research grant (VIL54489) from VILLUM FONDEN. Funding for the Sloan Digital Sky Survey IV has been provided by the Alfred P. Sloan Foundation, the U.S. Department of Energy Office of Science, and the Participating Institutions. SDSS acknowledges support and resources from the Center for High-Performance Computing at the University of Utah. The SDSS website is \url{www.sdss4.org}. SDSS is managed by the Astrophysical Research Consortium for the Participating Institutions of the SDSS Collaboration including the Brazilian Participation Group, the Carnegie Institution for Science, Carnegie Mellon University, Center for Astrophysics | Harvard \& Smithsonian (CfA), the Chilean Participation Group, the French Participation Group, Instituto de Astrofísica de Canarias, The Johns Hopkins University, Kavli Institute for the Physics and Mathematics of the Universe (IPMU) / University of Tokyo, the Korean Participation Group, Lawrence Berkeley National Laboratory, Leibniz Institut für Astrophysik Potsdam (AIP), Max-Planck-Institut für Astronomie (MPIA Heidelberg), Max-Planck-Institut für Astrophysik (MPA Garching), Max-Planck-Institut für Extraterrestrische Physik (MPE), National Astronomical Observatories of China, New Mexico State University, New York University, University of Notre Dame, Observatório Nacional / MCTI, The Ohio State University, Pennsylvania State University, Shanghai Astronomical Observatory, United Kingdom Participation Group, Universidad Nacional Autónoma de México, University of Arizona, University of Colorado Boulder, University of Oxford, University of Portsmouth, University of Utah, University of Virginia, University of Washington, University of Wisconsin, Vanderbilt University, and Yale University.

\section*{Data Availability}
The SDSS DR7 data are publicly available. The MaNGA data used are also publicly available at \url{https://www.sdss4.org/dr17/manga/manga-data/}. The dynamical masses and velocity dispersions are available from \citet{zhu_manga_2023}. The MPA-JHU catalogue is publicly available at \url{https://wwwmpa.mpa-garching.mpg.de/SDSS/DR7/}. The morphological parameters of SDSS galaxies are available from \citet{simard_catalog_2011}. The satellite classifications from \citet{yang_halo-based_2005, yang_galaxy_2007} are available at \url{https://gax.sjtu.edu.cn/data/Group.html}. 



\bibliographystyle{mnras}
\bibliography{articles} 



\appendix

\section{Properties Included in The Random Forest Regression}

We present a description of the various galaxy properties included in the random forest analysis for the MaNGA survey in Table~\ref{tab:manga_rf_parameters}.

\begin{table*}
    \centering
    \renewcommand{\arraystretch}{1.5}
  \caption{Parameters used for the random forest analysis using MaNGA data as shown in Figure~\ref{fig:manga_rf}.}
    \begin{tabular}{c|c}
    \hline
     Parameter &  Description \\
     \hline
      $\mathrm{M}_{\mathrm{*,spec,tot}}$ & total spectroscopic stellar mass from the global pyPipe3D catalouge \\
      $\mathrm{M}_{\mathrm{*,spec,tot,Re}}$ & total spectroscopic stellar mass within 1$\mathrm{R_e}$ from the global pyPipe3D catalogue \\
      $\mathrm{M}_{\mathrm{*,spec,sum,Re}}$ & total spectroscopic stellar mass within 1$\mathrm{R_e}$ computed by taking the sum of stellar masses per spaxel \\
       $\mathrm{M}_{\mathrm{*,phot}}$ & total photometric stellar mass from the global pyPipe3D catalouge \\ 
       $\Phi_{\mathrm{*,spec,tot}}$ & gravitational potential via $\mathrm{M}_{\mathrm{*,spec,tot}}/\mathrm{R_e}$ \\
       $\Phi_{\mathrm{*,spec,tot,Re}}$ & gravitational potential via $\mathrm{M}_{\mathrm{*,spec,tot,Re}}/\mathrm{R_e}$ \\
       $\Phi_{\mathrm{*,spec,sum,Re}}$ & gravitational potential via $\mathrm{M}_{\mathrm{*,spec,sum,Re}}/\mathrm{R_e}$ \\
       $\Phi_{\mathrm{*,phot}}$ & gravitational potential via $\mathrm{M}_{\mathrm{*,phot}}/\mathrm{R_e}$ \\
       $\mathrm{SFR}$ & integrated SFR using only SF spaxels taken from the global pyPipe3D catalouge \\
       $\mathrm{M}_{\mathrm{dyn,R_e}}$ & dynamical mass from \citealt{zhu_manga_2023} within $\mathrm{1R_e}$\\
       $\Phi_{\mathrm{dyn,R_e}}$ & dynamical gravitational potential via $\mathrm{M}_{\mathrm{dyn}}/\mathrm{R_e}$ \\
       $\mathrm{R_e}$ & effective radius \\
       $\mathrm{\sigma_{e}}$ & effective velocity dispersion from \citealt{zhu_manga_2023} within $\mathrm{1R_e}$ \\
       $\mathrm{Random}$ & random control variable \\
    \hline
    \end{tabular}
    \label{tab:manga_rf_parameters}
\end{table*}

\section{Reproducibility of and Comparison to Previous Results}\label{sec:app_reprod}

We compare our results with those of \citet{baker_stellar_2023} and \citet{sanchez-menguiano_stellar_2024}, both of which applied random forest regression to MaNGA data but reported opposite conclusions. By reproducing their sample selections and parameter sets, we can replicate their main results, allowing a direct comparison with our findings.

\citet{baker_stellar_2023} analysed a sample of roughly 1,000 galaxies, less than a quarter of the size of our dataset, and considered only spectroscopic stellar mass measurements from the pyPipe3D catalogue, without including photometric estimates. They concluded that gas-phase metallicity fundamentally depends on the spectroscopic stellar mass. In contrast, \citet{sanchez-menguiano_stellar_2024} used parameters exclusively from pyPipe3D measured at one effective radius, defining metallicity from a linear fit to the radial profile between $\rm 0.5,R_{\mathrm{e}}$ and $\rm 1.5,R_{\mathrm{e}}$, evaluated at $\rm 1.0,R_{\mathrm{e}}$. Their analysis therefore, omits the central $\rm 0.5,R_{\mathrm{e}}$ region, and they reported the baryonic gravitational potential derived from photometric stellar mass as the primary metallicity driver.

Our results agree with \citet{sanchez-menguiano_stellar_2024} that photometric stellar mass carries substantially more predictive power for gas-phase metallicity than its spectroscopic counterpart. This likely reflects the higher S/N of broadband photometric SED fits compared to spectroscopy, as well as the tendency of tree-based algorithms like random forest to favour lower-noise, higher-variance features \citep{kauffmann_host_2003, drory_comparing_2004, sanchez-menguiano_stellar_2024}. However, we differ from \citet{sanchez-menguiano_stellar_2024} in the relative importance of mass versus gravitational potential.

By examining feature importance as a function of galactocentric radius, we find that stellar mass dominates within the inner regions ($\rm \lesssim 0.7,R_{\mathrm{e}}$), while the baryonic gravitational potential becomes more important at larger radii ($\rm >0.9,R_{\mathrm{e}}$). This provides a natural explanation for the differences: because \citet{sanchez-menguiano_stellar_2024} excluded the central $\rm 0.5\mathrm{R_e}$ region (where we find stellar mass to be most influential), their analysis emphasised the gravitational potential. Our approach, which includes the galaxy centre, recovers a stronger mass dependence. 

These comparisons highlight that discrepancies between studies arise not only from differences in sample size and parameter choice, but also from the spatial regions over which metallicity is defined. As previous work has shown, metallicity correlations and gradients are highly sensitive to the spatial scale of the measurement \citep{belfiore_sdss_2017, belfiore_bathtub_2019, maiolino_re_2019, sharda_interplay_2024, garcia_metallicity_2025}.

\section{Differential Measurement Uncertainty Test}\label{sec:app_diff_unc}

A potential limitation of our core findings is that differences in measurement uncertainty among the input variables may bias the partial correlation coefficients and random forest regression, potentially affecting the inferred importance of each quantity. For example, if $\mathrm{M_{*,phot}}$ has much lower uncertainties than other mass measurements, specifically $\mathrm{M_{dyn,R_e}}$, then a weaker correlation would be found between $\mathrm{12+log(O/H)}$ and $\mathrm{M_{dyn,R_e}}$, without this being a true representation of the actual correlation between parameters. This is specifically interesting to investigate for the case of the dynamical mass, as we previously concluded that its low correlation with the gas-phase metallicity might be because it is only measured within $\mathrm{1R_e}$, as opposed to the entire galaxy, as for $\mathrm{M_{*,phot}}$. 

\begin{figure}
    \centering
    \includegraphics[width=\linewidth]{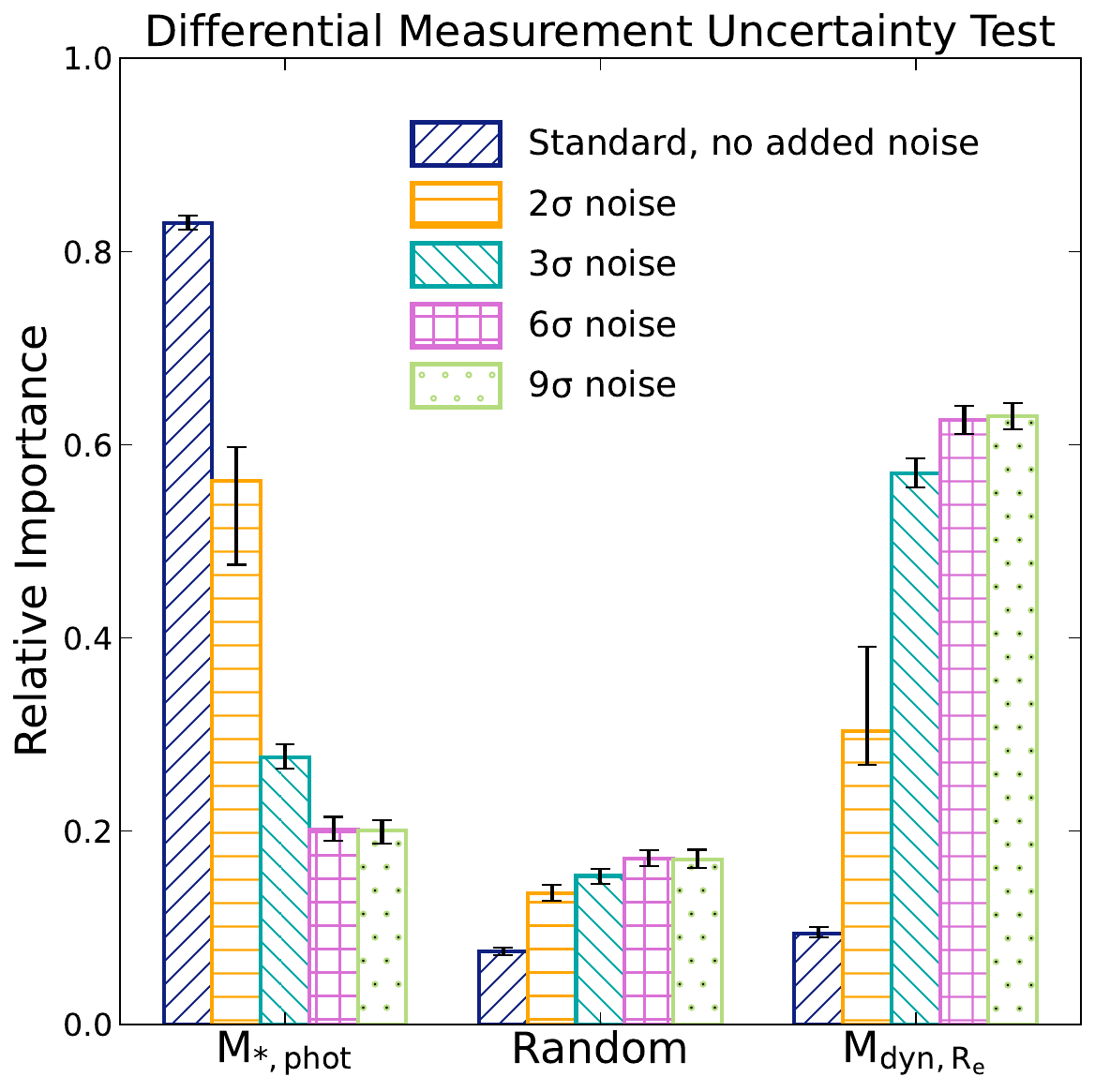}
    \caption{Random forest regression for the spectro-photometric stellar mass $\mathrm{M_{*,phot}}$, the dynamical mass $\mathrm{M_{dyn,R_e}}$, and a uniform random variable in determining the gas-phase metallicity. The four cases outlined are: standard case, and after addition of $\mathrm{2 \sigma_{M_{dyn,R_e}}}$, $\mathrm{3 \sigma_{M_{dyn,R_e}}}$, $\mathrm{6 \sigma_{M_{dyn,R_e}}}$, and $\mathrm{9 \sigma_{M_{dyn,R_e}}}$ worth of Gaussian random noise to $\mathrm{M_{*,phot}}$.}
    \label{fig:rf_diff_test}
\end{figure}

To look into this in detail, we additionally perform a differential measurement uncertainty test by artificially adding Gaussian random noise in units of the uncertainty of $\mathrm{M_{dyn,R_e}}$ \citep[$0.24$ dex, see][]{campbell_knowing_2017} to $\mathrm{M_{*,phot}}$ and re-evaluate their importances in driving the gas-phase metallicity via a random forest regression. Figure~\ref{fig:rf_diff_test} shows the relative importances of $\mathrm{M_{*,phot}}$, uniform random variable, and $\mathrm{M_{dyn,R_e}}$ in driving the gas-phase metallicity for the cases of no noise and different noise levels being added to the spectro-photometric stellar mass. We find that by adding $\mathrm{\geq 3\sigma_{M_{dyn,R_e}}}$ Gaussian random noise to $\mathrm{M_{*,phot}}$ the importance of $\mathrm{M_{*,phot}}$ starts to diminish and $\mathrm{M_{dyn,R_e}}$ becomes the main driver of the metallicity. However, the importance of $\mathrm{M_{dyn,R_e}}$ doesn't fully reach the high importance of $\mathrm{M_{*,phot}}$ even with an additional noise of $\mathrm{9 \sigma_{M_{dyn,R_e}}}$. Additionally, adding $\mathrm{9 \sigma_{M_{dyn,R_e}}}$ levels of noise results in $\mathrm{M_{*,phot}}$ being as (un-)important as a random variable.

The flip with a $\mathrm{3\sigma_{M_{dyn}}}$ noise suggests that differences in precision, not only intrinsic physics, helped elevate $\mathrm{M_{*,phot}}$ as the primary driver of gas-phase metallicity in our initial analysis (see Figure~\ref{fig:manga_rf}). Once that advantage is removed, $\mathrm{M_{dyn,R_e}}$ becomes the primary predictor, implying its intrinsic predictive power is at least comparable. However, this does not prove causality, but it strengthens the case that $\mathrm{M_{dyn,R_e}}$ carries a real significance that was previously underestimated due to larger measurement uncertainty.

\section{Combined Effect of Stellar Mass and Effective Radius}

We include an additional random forest analysis by conducting a similar investigation into the dependence of $\mathrm{M_{*,phot}/R_e^{\alpha}}$ as \citet{sanchez-menguiano_stellar_2024} has done. Essentially, we investigate the relation between the gas-phase metallicity and $\mathrm{M_{*,phot}/R_e^{\alpha}}$ for different $\mathrm{\alpha \in [0.1, 0.2, ..., 2.0]}$. For $\mathrm{\alpha=0}$ we establish the stellar mass, for $\mathrm{\alpha=1.0}$ we have the gravitational potential $\mathrm{\Phi_*}$, and for $\mathrm{\alpha=2}$ we get the mass surface density $\mathrm{\Sigma_*}$. \citet{sanchez-menguiano_stellar_2024} reasoning behind this exploration is that the correlation between the gas-phase metallicity and both stellar mass and galaxy size has been previously found \citep{ellison_clues_2008, sanchez_almeida_origin_2018, sanchez_mass-metallicity_2017}. Similarly, \citet{ma_revisiting_2024} recently demonstrated that combining stellar mass and size, through $\mathrm{M_*/R_e^{\alpha}}$ with $\alpha\sim0.6-1$, provides a much tighter correlation with gas-phase metallicity than stellar mass alone, in both MaNGA observations and TNG50 simulations. Within our analysis above, we include $\mathrm{\Phi_*=M_*/R_e}$, whose contribution to the metallicity could be attributed to a combined effect from stellar mass and galaxy size.  

\begin{figure}
    \centering
    \includegraphics[width=\linewidth]{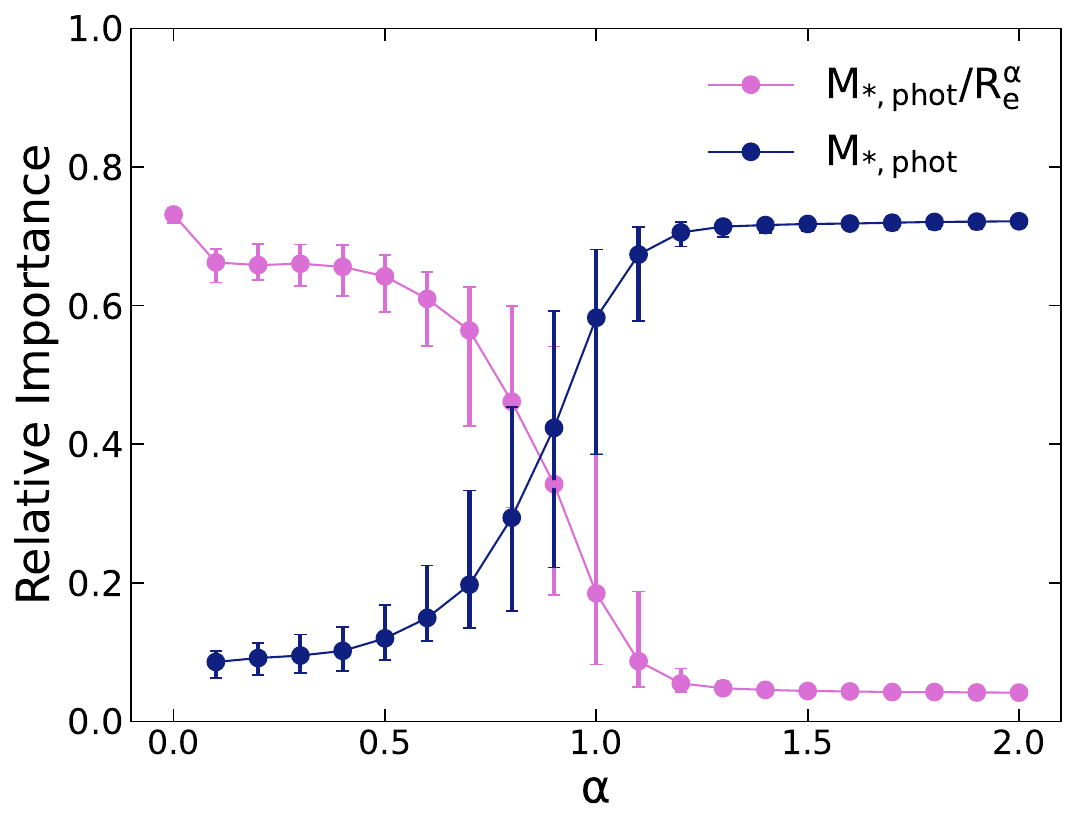}
    \includegraphics[width=\linewidth]{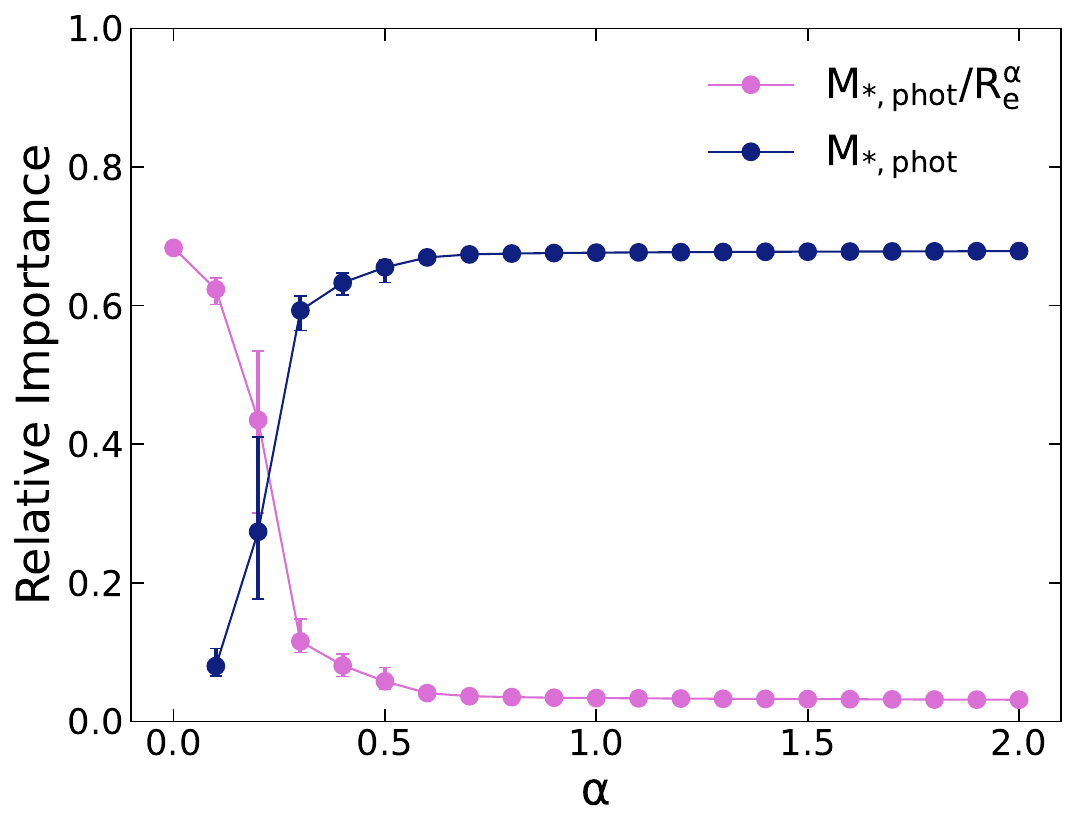}
    \caption{Importances of the photometric stellar mass $\mathrm{M_{*,phot}}$ (blue) and the ratio $\mathrm{M_{*,phot}/R_e^{\alpha}}$ (magenta) as we iterate through different $\alpha$. The error bars represent the 16th and 84th percentiles of 100 bootstrap random samples. For $\mathrm{\alpha = 0}$, we have the stellar mass and only include its importance as part of the blue data points. For $\mathrm{\alpha=1}$ and $\mathrm{\alpha=2}$ we get the gravitational potential $\mathrm{\Phi_{*,phot}}$ and surface mass density 
$\mathrm{\Sigma_{*,phot}}$ respectively. We present results for MaNGA (top figure) and SDSS (bottom figure) and find that the stellar mass overwhelmingly remains the key driver of the metallicity, but that for smaller $\mathrm{\alpha}$  there is a significant dependence present.}
    \label{fig:rf_manga_M_Re}
\end{figure}

We present our results for the MaNGA survey in Figure~\ref{fig:rf_manga_M_Re}  (top figure), where the magenta data points indicate the importance of $\mathrm{M_{*,phot}/R_e^{\alpha}}$ and the blue points are the stellar mass. For an $\mathrm{\alpha=0}$, we only include the stellar mass once in the random forest; thus, there is no blue datapoint for this. Besides the already confirmed results of the stellar mass being the main driver of the gas-phase metallicity, we also find that there is a significant importance of $\mathrm{M_{*,phot}/R_e^{\alpha}}$ for $\mathrm{\alpha}$ of roughly 0.1 to 0.8, when it dominates over the stellar mass. This is comparable to \citet{sanchez-menguiano_stellar_2024} results of $\mathrm{\alpha=0.6}$. For $\mathrm{\alpha>0.8}$, we confirm that $\mathrm{M_*}$ remains the dominant driver of the gas-phase metallicities. 

We repeat the same analysis using the SDSS sample, which we show at the bottom of Figure~\ref{fig:rf_manga_M_Re}. Here, the dependence switches to $\mathrm{M_*}$ at even lower $\mathrm{\alpha}$ values of roughly $\mathrm{\alpha > 0.2}$. The SDSS sample confirms that the spectro-photometric stellar mass predominantly dominates the gas-phase metallicity.

\section{Partial Correlation Coefficients Analysis for SDSS}
\begin{figure}
    \centering
    \includegraphics[width=\linewidth]{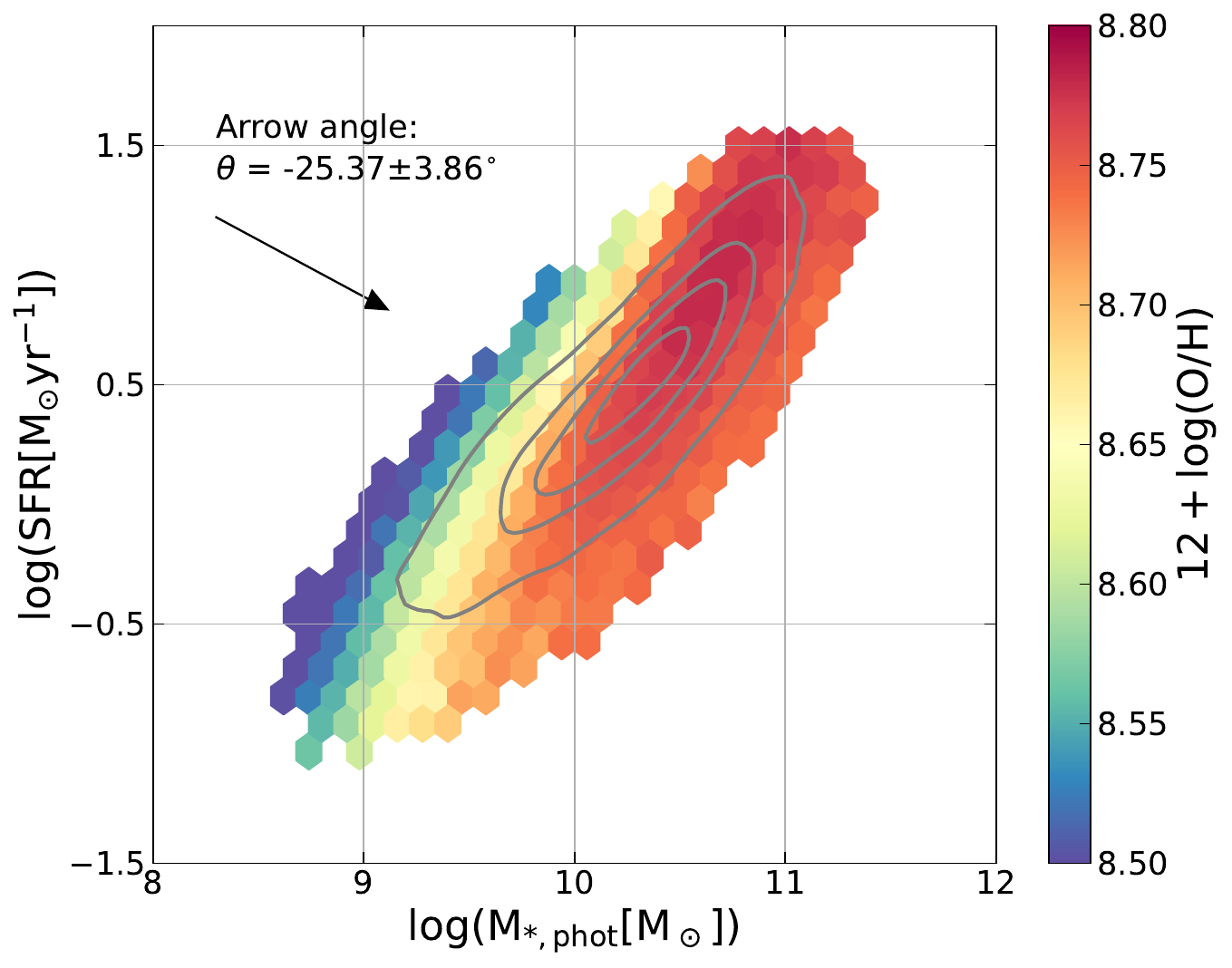}
    \caption{2D histogram of the photometric stellar mass versus SFR for the SDSS survey. The hexagonal bins are colour-coded by the gas-phase metallicity, and the grey contours indicate the galaxy distribution, with the outermost contour enclosing $90 \%$ of the data. The arrow gives the direction of the steepest gradient of the metallicity. The resulting angle is very similar to that obtained for the MaNGA survey selected for $\mathrm{\log(M_{*,phot})<10.5}$, confirming the presence of an FMR in the local universe.}
    \label{fig:hexbin_sdss}
\end{figure}

Here, we focus on the results of our PCC analysis via the SDSS sample as shown in Figure~\ref{fig:hexbin_sdss}.  Again, we see indications of the FMR in both the colour-shading and the arrow angle, confirming that the gas-phase metallicity has a stronger dependence on the stellar mass and a secondary inverse dependence on the SFR. 

\begin{figure}
    \centering
    \includegraphics[width=\linewidth]{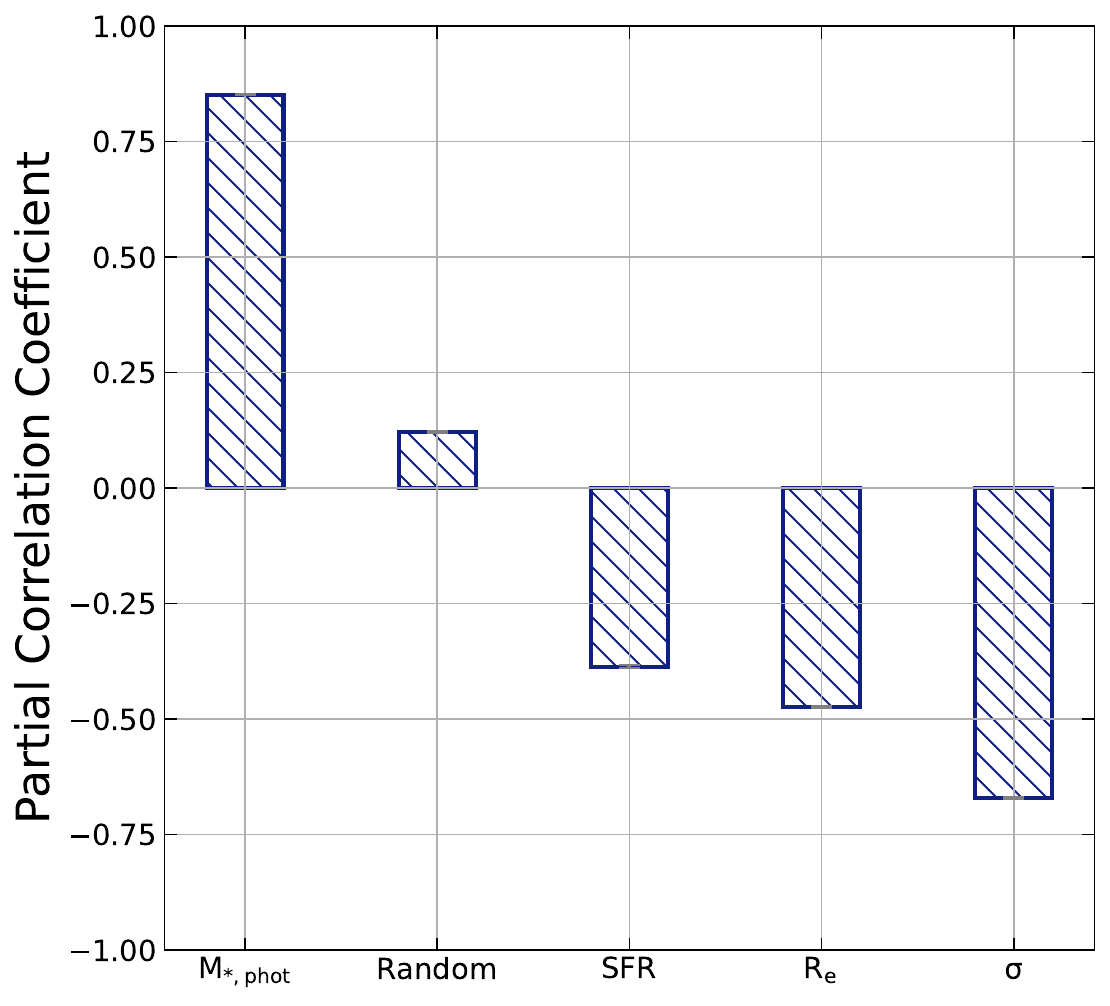}
    \caption{Partial Correlation Coefficients (PCC) between the gas-phase metallicity and the photometric stellar mass $\mathrm{M_{*,phot}}$, SFR, effective radius $\mathrm{R_e}$, velocity dispersion from Balmer emission lines $\mathrm{\sigma}$, and a uniform random variable (Random). Error bars are obtained by bootstrap random sampling 100 times. As for the random forest, PCCs indicate that the stellar mass is the main driver of metallicity, with a secondary dependence on the SFR. As already seen for the MaNGA data, the PCCs show a much stronger contribution from the other parameters than the random forest.}
    \label{fig:pcc_sdss}
\end{figure}

Figure~\ref{fig:pcc_sdss} presents the PCC results for the SDSS sample. As we do not have any measurements from dynamical modelling, we include only four different parameters in our analysis. As also seen in the random forest results (see Figure~\ref{fig:rf_sdss}), our results reveal a strong dependence on the stellar mass as well as the SFR. Furthermore, the effective radius $\mathrm{R_e}$ and velocity dispersion $\sigma$ show a much stronger correlation than for the MaNGA sample or the random forest results. 

\section{Central versus Satellite Galaxies with SDSS}

\begin{figure}
    \centering
    \includegraphics[width=\linewidth]{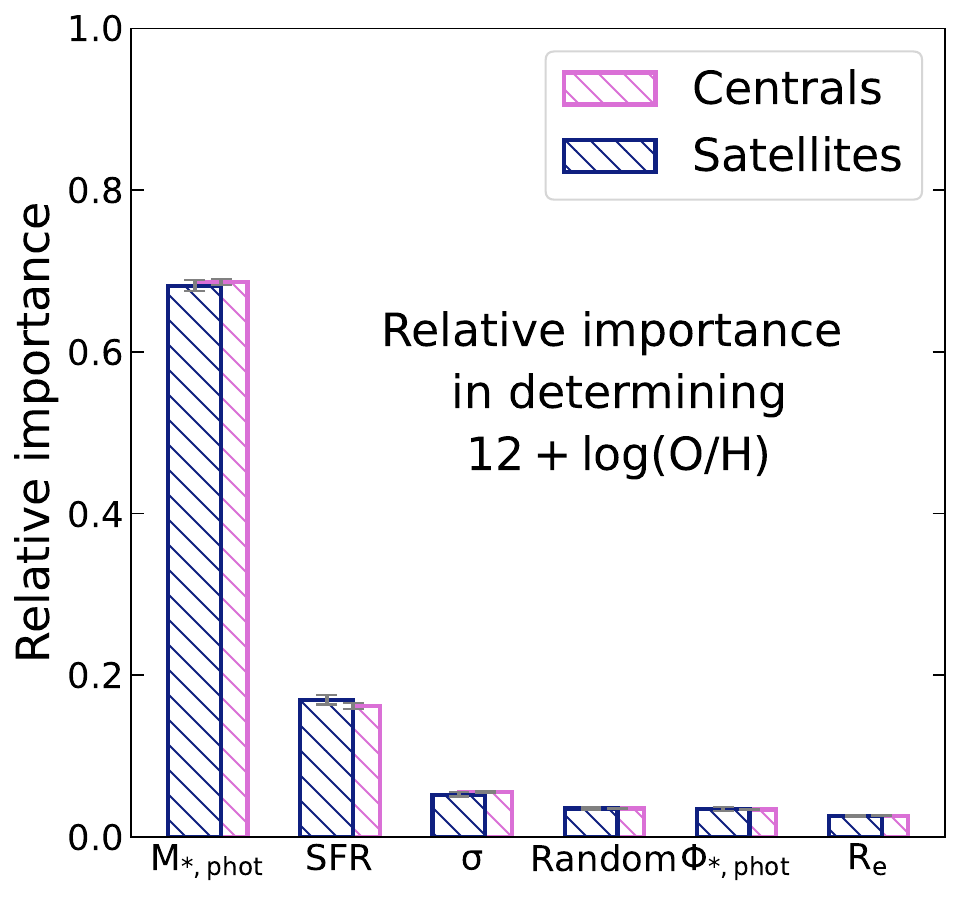}
    \caption{Random forest analysis for our SDSS sample as done in Figure~\ref{fig:rf_sdss} while separating central (pink) and satellite (blue) galaxies. The error bars represent the 16th and 84th percentiles from 100 bootstrap random samples. Our results do not indicate any differences in key drivers of the gas-phase metallicity between satellite and central galaxies.}
    \label{fig:rf_sdss_central_vs_sat}
\end{figure}

We use central satellite classifications from \citet{yang_halo-based_2005, yang_galaxy_2007} to identify satellite galaxies among our sample. Figure~\ref{fig:rf_sdss_central_vs_sat} shows the random forest analysis as done for Figure~\ref{fig:rf_sdss} while separating central and satellite galaxies. We do not find any difference in the results between central and satellite galaxies, as has previously been found for the stellar metallicity by \citet{baker_different_2024}.

\bsp	
\label{lastpage}
\end{document}